\pdfoutput=1
\PassOptionsToPackage{table}{xcolor}
\documentclass[11pt]{article}

\usepackage[final]{EMNLP2023}
\usepackage{times}
\usepackage{latexsym}
\usepackage[T1]{fontenc}
\usepackage[utf8]{inputenc}
\usepackage{microtype}
\usepackage{inconsolata}

\usepackage{xspace}
\usepackage{graphicx}
\usepackage{multirow}
\usepackage{amsfonts}

\usepackage{comment}
\usepackage{amsmath}
\usepackage{pifont}
\usepackage{tabularx}
\usepackage{caption, booktabs}
\usepackage{algorithm}
\usepackage{algpseudocode}
\usepackage{amssymb}
\usepackage{setspace}
\usepackage{bbm}
\usepackage{arydshln}

\makeatletter
\newcommand\footnoteref[1]{\protected@xdef\@thefnmark{\ref{#1}}\@footnotemark}
\makeatother
\newcolumntype{P}[1]{>{\centering\arraybackslash}p{#1}}

\usepackage{xspace}
\newcommand{\ours}[0]{\textsc{ArchCode}\xspace}
\usepackage{graphicx}
\usepackage{adjustbox}
\usepackage{subfig}
\makeatletter
\newcommand{\thickhline}{
    \noalign {\ifnum 0=`}\fi \hrule height 1pt
    \futurelet \reserved@a \@xhline
}
\newcolumntype{"}{@{\hskip\tabcolsep\vrule width 1pt\hskip\tabcolsep}}
\makeatother
\usepackage{mathabx}
\usepackage{amsfonts}

\makeatletter
\newcommand*{\blackleq}{
  \mathrel{
    \mathpalette\@blackleq{}
  }
}
\newcommand*{\@blackleq}[2]{
  \vcenter{
    \m@th
    \setbox0=\hbox{$#1\mkern3mu$}
    \setbox2=\hbox{$#1\vcenter{}$}
    \setbox4=\hbox{\raisebox{-\ht2}[.2pt][.2pt]{$#1-$}}
    \hbox{$#1\blacktriangleleft$}
    \nointerlineskip
    \kern\wd0 
    \copy4 
  }
}
\makeatother
\usepackage{dsfont}

\usepackage{multirow}
\usepackage{kotex} %

\usepackage{mathtools}

\definecolor{my_blue}{RGB}{0,112,192}
\newcommand{\cmark}{\selectcolormodel{cmy} \textcolor{green}{\ding{51}}}
\newcommand{\xmark}{\selectcolormodel{cmy} \textcolor{red}{\ding{55}}}

\usepackage{placeins}

\usepackage{tikz} 
\newcommand*\numcircledtikz[1]{\tikz[baseline=(char.base)]{
            \node[shape=circle,draw,inner sep=0.05pt] (char) {#1};}}

\title{\ours: Incorporating Software Requirements in\\ Code Generation with Large Language Models}

\author{
Hojae Han$^{\spadesuit}$$^{\diamond}$ $ $ Jaejin Kim$^{\spadesuit}$$^{\diamond}$ $ $ Jaeseok Yoo$^{\spadesuit}$ $ $ Youngwon Lee$^{\spadesuit}$$^{\diamond}$ $ $ Seung-won Hwang$^{\spadesuit}$$^{\diamond}$\thanks{~~~Corresponding author.}\\
$^{\spadesuit}$Seoul National University, $^{\diamond}$SNU-LG AI Research Center\\
\texttt{\{stovecat,jaejin.kim,jaeseok2.yoo,ludaya,seungwonh\}@snu.ac.kr}
}

\begin{document}
\maketitle

\begin{abstract}
This paper aims to extend the code generation capability of large language models (LLMs) to automatically manage comprehensive software requirements from given textual descriptions.
Such requirements include both functional (i.e. achieving expected behavior for inputs) and non-functional (e.g., time/space performance, robustness, maintainability) requirements. 
However, textual descriptions can either express requirements verbosely or may even omit some of them. 
We introduce \ours, a novel framework that leverages in-context learning to organize requirements observed in descriptions and to extrapolate unexpressed requirements from them. 
\ours~generates requirements from given descriptions, conditioning them to produce code snippets and test cases.
Each test case is tailored to one of the requirements, allowing for the ranking of code snippets based on the compliance of their execution results with the requirements.
Public benchmarks show that \ours~enhances to satisfy functional requirements, significantly improving Pass@$k$ scores. 
Furthermore, we introduce HumanEval-NFR, the first evaluation of LLMs' non-functional requirements in code generation, demonstrating \ours's superiority over baseline methods.
The implementation of \ours~and the HumanEval-NFR benchmark are both publicly accessible.\footnote{\url{https://github.com/ldilab/ArchCode}}
\end{abstract}

\section{Introduction}
\label{intro}
Recent advancements in large language models (LLMs) have significantly improved code generation capabilities~\cite{codex,li2022competition,openai2023gpt4}.
Although the primary goal for LLMs in this domain is to generate functionally correct code based on textual descriptions~\cite{DBLP:conf/nips/HendrycksBKMAGB21,DBLP:journals/corr/abs-2108-07732,codex,li2022competition}, real-world software development encompasses more than just functionality.

\begin{figure}[t]
\centering
\hspace*{-0.3cm}
\begin{tabular}{cc}
\subfloat[Existing Approaches. \label{fig:diff_a}]{%
      \includegraphics[width=0.5\linewidth]{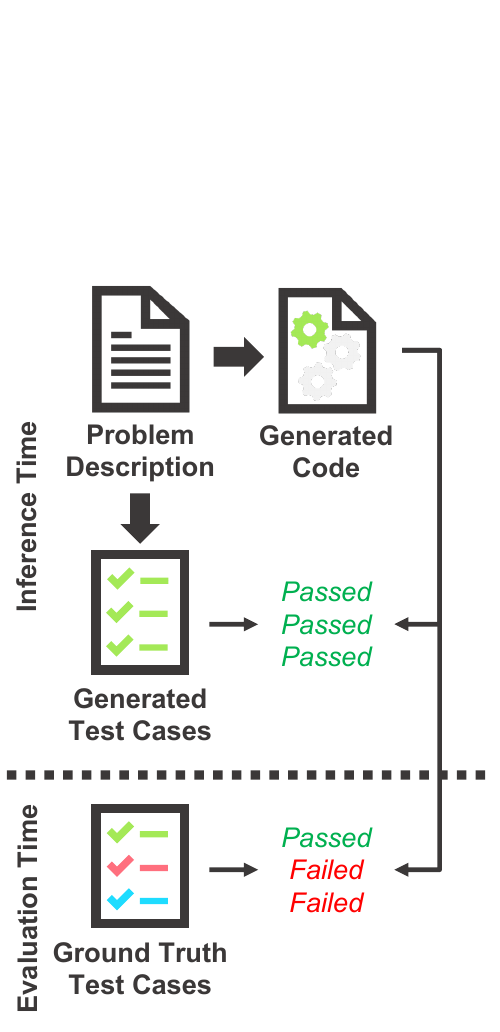}}
\hspace*{-0.3cm}
&
\subfloat[\ours~(this work). \label{fig:diff_b}]{%
      \includegraphics[width=0.5\linewidth]{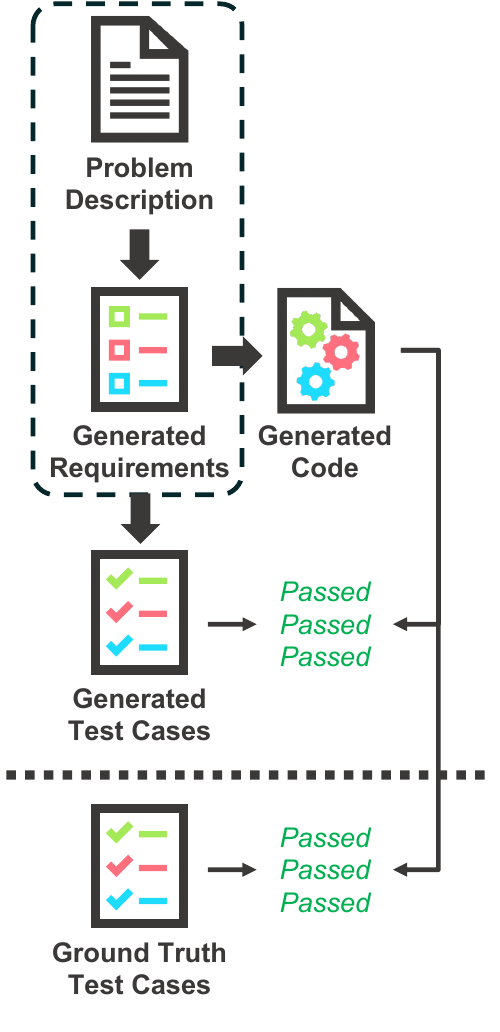}}\\
\end{tabular}
\caption{The \ours~framework infers software requirements of correct code solution for a given textual description, then conditions them to generate code, as well as test cases for verification.}
\label{fig:diff}
\end{figure}

\begin{figure*}[t]
\centering  
\includegraphics[width=0.9\linewidth]{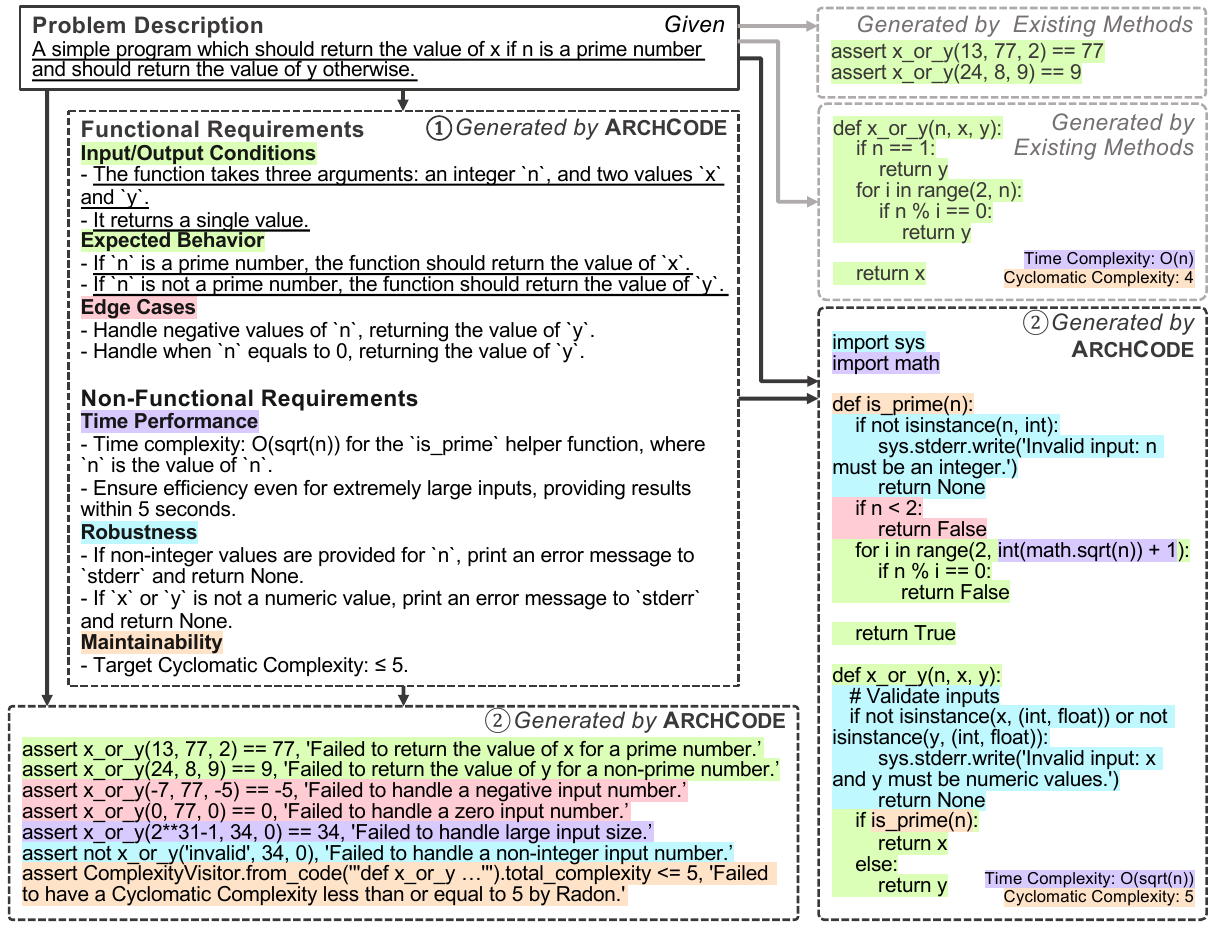}
\caption{An illustrative example of code and test case generation.
Existing approaches derive code and test cases directly from problem descriptions, often missing key requirements.
\ours, in contrast, \numcircledtikz{1} reformulates (underlined) and extrapolates (not underlined) requirements from these descriptions, then \numcircledtikz{2} generates code and test cases to meet them comprehensively.
Best viewed in color.}
    \label{fig:contribution_figure}
\end{figure*}

In software engineering, software requirements provide a detailed framework describing what a software system is intended to achieve~\cite{chung2012non}, divided into two categories~\cite{glinz2007non}:
\begin{itemize}
    \item \textbf{Functional Requirements} (FRs) dictate the behavior and functionality, e.g., input/output conditions, desired behavior of code, etc. 
    \item \textbf{Non-Functional Requirements} (NFRs) are attributes or constraints beyond functionality, e.g., time and space performance, robustness, maintainability, reliability, etc. 
\end{itemize}

{Despite the critical role of software requirements, considering these criteria has not been studied actively in previous code generation works, merely generating code directly from textual descriptions.}
However, textual descriptions might express requirements verbosely or even omit them. 
As illustrated in Figure~\ref{fig:diff_a} and~\ref{fig:contribution_figure} (upper right), this may result in code that neglects many desirable requirements. 
Code filtering based on generated test cases~\cite{li2022competition,chen2023codet,huang2023enhancing} shares the same problem, as test cases often fail to cover a broader range of requirements. 
Consequently, the generated code might exhibit unexpected behaviors for valid inputs, ignoring FRs. 
Similarly, overlooking NFRs can result in time/space inefficiencies, potential system failures, or challenges in maintenance. 
{Nevertheless, achieving conciseness in the textual descriptions of software requirements necessitates significant human effort~\cite{perry1992foundations,bass2003software}.}

{We introduce \ours, a novel framework that automatically incorporates software requirements from textual descriptions, then directs LLMs to align code and test case generation with those requirements,}
as illustrated in Figure~\ref{fig:diff_b}. 
{Specifically, \ours~leverages In-Context Learning (ICL;~\citealp{kojima2022large,10.5555/3618408.3619681,zhang2023automatic}) for adaptability, utilizing LLMs' extensive reasoning abilities to learn within context, thereby avoiding costly parameter updates.} 
For code generation, each in-context example comprises a triplet---a textual description, a list of software requirements (including both those expressed and unexpressed in the description), and corresponding code that satisfies all these requirements. 
For test case generation, we simply switch from code to test cases, each of which verifies a specific requirement.
{\ours~prepends in-context examples to test descriptions, guiding LLMs to: 1) reformulate explicit requirements in descriptions, 2) deduce implicit requirements from their parametric knowledge, 3) generate code that fulfills these requirements, and 4) produce test cases for verifying each requirement}, as shown in Figure~\ref{fig:contribution_figure}.

We integrate \ours~with \textit{WizardCoder}~\cite{luo2023wizardcoder} and GPT-3.5-Turbo~\cite{openai2022chatgpt}, and assess the performance on HumanEval~\cite{codex} and CodeContests~\cite{li2022competition}. 
The results confirm that \ours notably outperforms existing techniques in terms of the satisfaction of FRs---surpassing GPT-4's Pass@$1$ score on both benchmarks and  achieving new state-of-the-art on CodeContests. 
Moreover, we introduce HumanEval-NFR based on HumanEval, the first benchmark to evaluate NFRs alongside FRs, to confirm that \ours is also effective in pursuing NFRs.

Our main contributions are as follows:
\begin{itemize}
\item We propose \ours, a novel framework that leverages ICL to incorporate software requirements in code generation.
\item \ours with GPT-3.5-Turbo surpasses GPT-4's Pass@$1$ scores on both HumanEval and CodeContests by 4.81\%p and 10.45\%p, while requiring $50\times$ smaller number of test cases to be generated compared to existing methods.
\item We introduce HumanEval-NFR, the first code generation benchmark for NFR evaluation to confirm the effectiveness of \ours~for NFR satisfaction.
\end{itemize}

\section{Related Work}
\label{related}
Despite the fact that LLMs recently have shown impressive capabilities in code generation, the majority of evaluations have focused solely on functional requirements (FRs;~\citealp{DBLP:conf/nips/HendrycksBKMAGB21,DBLP:journals/corr/abs-2108-07732,codex,li2022competition}).

\begin{table}[t]
    \resizebox{0.475\textwidth}{!}{
    \begin{tabular}{lcc}
        \thickhline
         & \multirow{1}{*}{\small{FRs}} & \multirow{1}{*}{\small{NFRs}} \\
        \hline
        {Self-planning}\small{~\cite{jiang2023selfplanning}}  & { \cmark}      & { \xmark}      \\ 
        {\textsc{Brainstorm}}\small{~\cite{li2023think}}  & { \cmark}      & { \xmark}      \\ 
        {\textsc{Algo}\small{~\cite{zhang2023algo}}}  & { \cmark}      & { \xmark}      \\ 
        {\textsc{CodeRanker}}\small{~\cite{inala2022faultaware}}  & { \cmark}      & { \xmark}      \\ 
        {{\textsc{Lever}}\small{~\cite{ni2023lever}}}  & { \cmark}      & { \xmark}      \\ 
        {{Coder-Reviewer}}\small{~\cite{zhang2023coder}}  & { \cmark}      & { \xmark}      \\ 
        {AlphaCode}\small{~\cite{li2022competition}}           & { \cmark} & { \xmark} \\ 
        {{\textsc{MBR-Exec}}\small{~\cite{shi-etal-2022-natural}}}       & { \cmark} & { \xmark} \\
        {\textsc{CodeT}}\small{~\cite{chen2023codet}}       & { \cmark} & { \xmark} \\
        {\textsc{Reflexion}}\small{~\cite{shinn2023reflexion}}   & { \cmark} & { \xmark} \\
        {\textsc{MPSC}}\small{~\cite{huang2023enhancing}}   & { \cmark} & { \xmark} \\
        {\textit{WizardCoder}}\small{~\cite{luo2023wizardcoder}}  & { \cmark} & \phantom{ }$\triangle$ \\
        {\textsc{Pie}}\small{~\cite{madaan2023learning}}  & { \xmark} & \phantom{ }$\triangle$ \\
        {\textsc{TitanFuzz}}\small{~\cite{deng2023large}}  & { \xmark} & \phantom{ }$\triangle$ \\
        {\textsc{Fuzz4All}}\small{~\cite{xia2023universal}}  & { \xmark} & \phantom{ }$\triangle$ \\
        \rowcolor{gray!10}
        {\ours}\small{ (this work)}                & { \cmark} & { \cmark} \\
        \thickhline
    \end{tabular}}
    \caption{\ours~is a novel code and test case generation framework that pursues the satisfaction of both FRs and NFRs. {In NFRs column, $\triangle$ denotes that only one or two NFRs were addressed in those works, whereas \ours~addresses four different NFR categories, marked as\cmark.}}
    \label{tab:contribution}
\end{table}

\paragraph{Solely Targeting Functional Requirements}
Early research such as~\citet{feng-etal-2020-codebert},~\citet{codex},~\citet{brown2020language},  and~\citet{li2022competition} 
directly generates code from natural language descriptions, which may not fully capture all software requirements due to their vagueness or imperfections. 
Later studies~\cite{jiang2023selfplanning,li2023think,zhang2023algo} have targeted to better capture functional requirements by generating code-like outlines via in-context learning (ICL;~\citealp{kojima2022large,10.5555/3618408.3619681,zhang2023automatic}).
More recent methods enhance FR satisfaction through self-verification of the generated code: 
On one hand, \textit{code filtering} utilizes `over-generate-then-filter' strategies, where the filtering can be achieved either by predicting functional correctness {without code execution~\cite{inala2022faultaware,ni2023lever,zhang2023coder}, or execution with given~\cite{shi-etal-2022-learning} or} generated test cases~\cite{li2022competition,chen2023codet,huang2023enhancing}.
On the other hand, \textit{code refinement} iteratively reflects and refines the generated code to improve its functionality via execution results of current version of code with generated test cases~\cite{shinn2023reflexion}.
\paragraph{Targeting Narrow Scope of Non-Functional Requirements}
While much research covers FRs, few studies have addressed specific attributes of non-functional requirements (NFRs) such as reliability and robustness~\cite{deng2023large,xia2023universal}, or time/space performance~\cite{madaan2023learning,luo2023wizardcoder}.

\paragraph{Our Distinction}
Table~\ref{tab:contribution} summarizes the distinction of our framework compared with existing code generation approaches. 
{To the best of our knowledge, \ours~is the first study that employs ICL to systematically extract and interpret software requirements from descriptions, ensuring the generated code and test cases closely aligns with these requirements. }
In addition, we introduce HumanEval-NFR, a variant version of the HumanEval~\cite{codex} benchmark that can assess the fulfillment of NFRs.

\begin{figure*}[htb!]
\centering  
\includegraphics[width=1.0\linewidth]{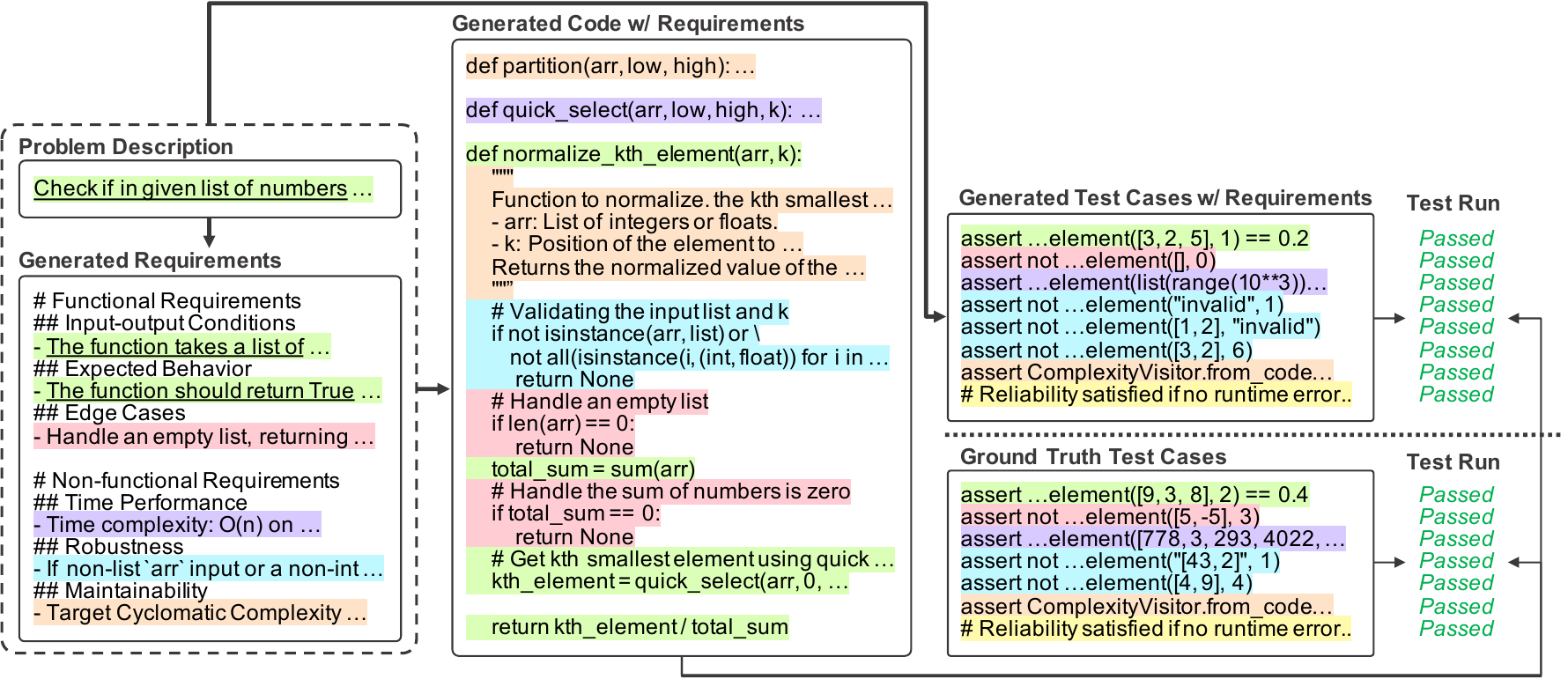}
\caption{The overview of the \ours~framework. Each color represents the subtype of software requirements. Underlined requirements are expressed in problem descriptions, whereas other requirements are inferred from descriptions by LLMs' parametric knowledge.
Best viewed in color.}
\label{fig:overview}
\end{figure*}

\section{The \ours~Framework}
We propose \ours, a novel code generation framework that employs In-Context Learning (ICL) to LLMs, incorporating software requirements in code generation.  %
As shown in Figure~\ref{fig:overview}, \ours~delineates software requirements from textual descriptions, generates code, then verifies it using custom test cases.\footnote{{More details on input/output formats, in-context examples and hyperparameters are provided in Appendix~\ref{impl_detail}.} }

Formally, given a problem space $\mathcal{P}$ and a code space $\mathcal{C}$, the code generation task is to build a function $\mathcal{F}$ : $\mathcal{P} \rightarrow \mathcal{C}$ that maps each textual problem description $p \in \mathcal{P}$ into its corresponding code implementation $c \in \mathcal{C}$. 
\ours~decomposes $\mathcal{F}$ by $\mathcal{F}=g \circ f$. 
$g$ : $\mathcal{P} \rightarrow \mathcal{P} \times \mathcal{R}$ maps problems to problem-requirements pairs, and
$f$ : $\mathcal{P} \times \mathcal{R} \rightarrow \mathcal{C}$ generates code from the problem-requirements pairs, where
$\mathcal{R}$ is a space of software requirements. 
The test case generation function $\mathcal{H}$ : $\mathcal{P} \rightarrow \mathcal{T}$ is also decomposed by \ours~into $\mathcal{H}= g \circ h$, where $\mathcal{T}$ is the space of test cases and $h$ : $\mathcal{P} \times \mathcal{R} \rightarrow \mathcal{T}$.

\subsection{Delineating Software Requirements}
\label{subsec:software_requirements}
\ours leverages ICL to let an LLM %
generate a set of software requirements, either reformulated from a given textual description, or extrapolated from the description by the LLM's learnt parametric knowledge.
Formally, given in-context examples of description-requirements pairs and the target description $p$, the LLM returns the list of software requirements 
\begin{align}
\label{eq:r}
    \hat{\textbf{r}} = g([p_1', \textbf{r}_1']; [p_2', \textbf{r}_2']; ...; [p]), 
\end{align}
where $p_i'$ and $\textbf{r}_i'=[r_i^1, r_i^2, ... ]$  is the description and its corresponding requirement list of $i$-th example pair, each $r_i^j$ in $\textbf{r}_i'$ is a requirement, 
and $\hat{\textbf{r}}$ is the list of generated requirements. 

Specifically, based on the established classifications by~\citet{glinz2007non}, we further break down FRs and NFRs into distinct categories that our study focuses on.

\paragraph{Target FRs}
Our approach narrows down FRs into three subtypes:
\begin{itemize}
    \item \textbf{Input/Output Conditions}: 
    Analogous to the preconditions and the postconditions in \textit{Design by Contract}~\cite{meyer1992applying}, these define the desired functionality of the code by specifying valid inputs and expected outputs. 
    \item \textbf{Expected Behavior}: Along with Input/Output Conditions, it explains the functionality of the target code by reformulating the description into a series of operations for valid inputs that are applicable in most general scenarios.
    \item \textbf{Edge Cases}: While this term generally comprises an array of corner cases, we restrict the scope to only consider valid inputs that necessitate distinct treatment. These include, for example, processing an empty list when the valid input type is a list, or considering `0' for non-negative integer inputs.
\end{itemize}

\paragraph{Target NFRs}
\ours~considers NFRs that are both pivotal in real-world applications and feasible for assessment either through code execution or using existing metrics. 
\begin{itemize}
    \item \textbf{Time Performance}: Pertains to time-centric aspects like algorithmic time complexity or stipulated timeout conditions. 
    \item \textbf{Robustness}: Ensures that code is resilient to invalid inputs~\cite{mcconnell2004code}. 
    For instance, a function designed for integer addition must prevent unforeseen or undesirable outcomes from the `+' operation, like mistakenly returning the concatenated string when given two strings.
    \item \textbf{Maintainability}: Considers factors that contribute to the ease of maintenance, such as 
    {reducing code complexity 
    via code modularization~\cite{magel1982applying}, measured by cyclomatic complexity~\cite{mccabe1976complexity}. }
    \item \textbf{Reliability}: Ensures that the code can handle errors gracefully, without causing system failures, thereby increasing the mean time between failures~\cite{mcconnell2004code}. 
\end{itemize}

\subsection{Requirements-aware Generation}
Upon obtaining software requirements $\hat{\textbf{r}}$, \ours conditions $\hat{\textbf{r}}$ with the given description $p$ to generate code samples and test cases.\footnote{For the reliability category, we uniquely assess code reliability by checking for runtime errors with various test cases, instead of generating specific ones.}
Specifically, \ours~generates code $\hat{c}$ and test cases $\hat{\textbf{t}}$ in a parallel manner: 
\begin{align}
    \label{eq:codegen}
    \hat{c} = & f([p_1', \textbf{r}_1', c_1']; ...; [p, \hat{\textbf{r}}]), \nonumber\\
    \hat{\textbf{t}} = & h([p_1', \textbf{r}_1', \textbf{t}_1']; ...; [p, \hat{\textbf{r}}]), 
\end{align}
where $c_i'$ and $\textbf{t}_i'=[t_i^1, t_i^2, ...]$ are the code and the list of test cases of $i$-th example, and each $t_i^j$ in $\textbf{t}_i'$ is a test case corresponding to $r_i^j $ in $\hat{\textbf{r}}$.\footnote{For an intuitive explanation, we describe how a single test case is tailored to a requirement. However, in real implementation, \ours~utilizes multiple generated test cases to confirm each requirement, as explained in Appendix~\ref{appendix:analysis_of_gen_tcs}.} 
We choose this parallel generation due to the potential pitfalls when these processes condition each other. We further discuss such pitfalls in Section~\ref{TDD}.

\begin{table*}[htb!]
    \centering
    \begin{tabular}{lccccccccccccccc}
        \thickhline
         & 
         \multicolumn{2}{c}{\multirow{1.25}{*}{All}} & 
         \multicolumn{2}{c}{\multirow{1.25}{*}{Time Perf.}} &
         \multicolumn{2}{c}{\multirow{1.25}{*}{Robustness}} &
         \multicolumn{2}{c}{\multirow{1.25}{*}{Maintainability}} &
         \multicolumn{2}{c}{\multirow{1.25}{*}{Reliability}} 
         \\
        \cmidrule(r){2-3}\cmidrule(r){4-5}\cmidrule(r){6-7}\cmidrule(r){8-9}\cmidrule(r){10-11}
         \multirow{-1.25}{*}{Pass@$k$} 
         & \multirow{-1.25}{*}{$k$=1} & \multirow{-1.25}{*}{5} &
           \multirow{-1.25}{*}{$k$=1} & \multirow{-1.25}{*}{5} &
           \multirow{-1.25}{*}{$k$=1} & \multirow{-1.25}{*}{5} &
           \multirow{-1.25}{*}{$k$=1} & \multirow{-1.25}{*}{5} &
           \multirow{-1.25}{*}{$k$=1} & \multirow{-1.25}{*}{5} 
           \\\hline

        GPT-3.5-Turbo & 
            {\small \phantom{0}2.62} & {\small 10.03} & 
            {\small 53.48} & {\small 65.75} & 
            {\small \phantom{0}4.21} & {\small 14.55} & 
            {\small 53.23} & {\small \underline{68.38}} & 
            {\small 20.98} & {\small 36.72} & 
            \\
        \phantom{ }\phantom{ }+\phantom{ }CoT & 
            {\small \phantom{0}\underline{5.00}} & {\small \underline{12.08}} & 
            {\small  50.00} & {\small  66.03} & 
            {\small \phantom{0}\underline{7.32}} & {\small \underline{17.67}} & 
            {\small 44.33} & {\small 62.00} & 
            {\small 45.49} & {\small 66.83} & 
            \\
        {\textsc{CodeT}} & 
            {\small \phantom{0}3.03} & {\small 10.03} & 
            {\small \underline{58.50}} & {\small \underline{67.31}} & 
            {\small \phantom{0}4.61} & {\small 14.52} & 
            {\small \textbf{57.80}} & {\small \textbf{68.50}} & 
            {\small \underline{47.62}} & {\small \textbf{74.90}} & 
            \\
        \rowcolor{gray!10} \ours & 
            {\small \textbf{25.19}} & {\small \textbf{27.33}} & 
            {\small \textbf{62.86}} & {\small \textbf{69.70}} & 
            {\small \textbf{40.86}} & {\small \textbf{42.72}} & 
            {\small \underline{56.43}} & {\small 62.23} & 
            {\small \textbf{68.53}} & {\small \underline{74.67}} & 
            \\

        \thickhline
    \end{tabular}
    \caption{Experimental results on HumanEval-NFR. Each column states the evaluation category of NFRs. Boldface and underline denote the 1st and 2nd highest scores, respectively.  MPSC is omitted as it is publicly unavailable. %
    } 
    \label{tab:humaneval_nfr}
\end{table*}

\subsection{Pursuing Requirements Satisfaction}
To ensure the conformance of the generated code snippet $\hat{c}$ with the specified requirements $\hat{\textbf{r}}$, \ours~executes $\hat{c}$ against the generated test cases $\hat{\textbf{t}}$ tailored to one of the requirements in $\hat{\textbf{r}}$:
\begin{align}
    s = \texttt{EXEC}\left(\hat{c}, \hat{t}\right), 
\end{align}
where $s\in \{0,1\}$ is a binary result from a code execution function $\texttt{EXEC}$, and $\hat{t}$ is one of the generated test cases in $\hat{\textbf{t}}$, matching $\hat{r}$ in $\hat{\textbf{r}}$. 
To return the satisfactory code towards $\hat{\textbf{r}}$, \ours~conducts code filtering. 
To rank each code in relation to $\hat{\textbf{r}}$, our framework calculates a weighted sum of the scores $s$ from each $\hat{t}$, with the option to assign higher weights to preferred requirements.
Adjusting those weights to tailor the scoring process is discussed in more detail in Section~\ref{weighting}.

\section{Experiments}
{We evaluate \ours's effectiveness using three benchmarks, categorized into two types: 
1) A novel benchmark for assessing both FR and NFR satisfaction; 
2) Two public benchmarks aimed at FR evaluation, facilitating comparison of \ours~with existing baselines. 
For the former, we introduce HumanEval-NFR for comprehensive NFR assessment, overcoming the conventional focus on FR alone. 
For the latter, we explore two code modalities: 1) function-level and 2) competition-level code generation.}

\subsection{Experimental Setup}
We evaluate the effectiveness of \ours~on code generation with LLMs. 
Throughout the experiments, we used 
GPT-3.5-Turbo-16k~\cite{openai2022chatgpt} as the backbone LLMs for generating code, software requirements, test cases, etc.
More details can be found in Appendix~\ref{impl_detail}.

\paragraph{Evaluation Metrics}
{We mainly consider the widely used Pass@$k:=\mathbb{E}_{\text{Problems}} [1-\frac{\binom{n-c}k}{\binom{n}k}]$~\cite{codex} metric for evaluation, which is the unbiased estimator of the probability that the code generation system would have passed a problem if it were given $k$ chances to sample $c$ correct code snippets among $n$ samples.} %
Adhering to~\citet{chen2023codet}, when applying code filtering, we denote the existence of passed code among the $k$ filtered samples.\footnote{While this metric is sometimes referred to as $n$@$k$---the pass ratio of filtered $n$ samples from $k$---we avoid this notation as the interpretation of $k$ differs from that in Pass@$k$.}

\begin{table*}[t]
    \centering
    \begin{tabular}{lcccccccccc}
        \thickhline
         &&\multirow{2.25}{*}{Code} &  & \multicolumn{6}{c}{\multirow{1.25}{*}{Pass@$k$}} \\
         \multirow{1.25}{*}{Method} & \multirow{1.25}{*}{CoT} & \multirow{2}{*}{Filtering} & \multirow{1.25}{*}{NFRs} & \multicolumn{3}{c}{\multirow{1.3}{*}{HumanEval}} & \multicolumn{3}{c}{\multirow{1.3}{*}{CodeContests}} \\
        \cmidrule(r){5-7} \cmidrule(r){8-10}
         & & & & \multirow{-1.25}{*}{$k$=1} & \multirow{-1.25}{*}{2} & \multirow{-1.25}{*}{5}
         & \multirow{-1.25}{*}{$k$=1} & \multirow{-1.25}{*}{2} & \multirow{-1.25}{*}{5} \\\hline

        \textsc{CodeRanker}\textsuperscript{\textdagger} & 
            \xmark & \cmark & \xmark &
            32.3\phantom{0} & 
            - & 
            61.6\phantom{0} & - & - & -
            \\
        \textit{WizardCoder} 34B\textsuperscript{\textdagger} & 
            \xmark & \xmark & \phantom{ }$\triangle$ &
            73.2\phantom{0} & 
            - & 
            - & - & - & -
            \\
        GPT-4\textsuperscript{\ddag} & 
            \xmark & \xmark & \xmark &
            81.55 & 86.39 & \textbf{90.49} &
            \phantom{0}6.07 & \phantom{0}8.23 & 11.67 
            \\
            \hdashline

        GPT-3.5-Turbo & 
            \xmark & \xmark & \xmark &
            73.17 & 80.79 & 86.99 &
            \phantom{0}4.79 & \phantom{0}7.02 & 10.06
            \\
        \phantom{ }\phantom{ }+\phantom{ }CoT & 
            \cmark & \xmark & \xmark &
            72.99 & 79.58 & 83.95 &
            \phantom{0}5.82 & \phantom{0}8.57 & 13.53
            \\
        \textsc{Brainstorm}\textsuperscript{\textdagger} &
            \cmark & \xmark & \xmark &
            - & - & - & 
            \phantom{0}7.0\phantom{0} &
            - &
            14.7\phantom{0}
            \\
        {\textsc{Algo}\textsuperscript{\textdagger}} &
            \cmark & \xmark & \xmark &
            {-} & {-} & {-} & 
            {12.00} &
            {12.00} &
            {-}
            \\
        {\textsc{MBR-Exec}\textsuperscript{\ddag}} &
            \xmark & \cmark & \xmark & 
            {72.96} & {76.47} & {79.00} & 
            \phantom{0}{8.25} & \phantom{0}{8.87} & \phantom{0}{9.08} 
            \\
        \textsc{CodeT}\textsuperscript{\ddag} & 
            \xmark  & \cmark & \xmark &
            78.05 & 78.05 & 78.30 &
            \phantom{0}9.92 & 10.18 & 10.30
            \\        
        MPSC\textsuperscript{\textdagger} &
            \cmark & \cmark & \xmark & 
            \underline{85.37} & \underline{86.60} & 86.35 & 
            \underline{14.39} & \textbf{17.16} & \textbf{17.76} \\
        \rowcolor{gray!10}
        \ours (Ours)          & 
            \cmark & \cmark & \cmark &
            \textbf{86.36} & \textbf{88.62} & \underline{90.48} &
            \textbf{16.52} & \underline{16.67} & \underline{17.37} 
            \\
        \thickhline
    \end{tabular}
    \caption{Experimental results on HumanEval and CodeContests.
    Daggers\textsuperscript{\textdagger} denote the values are directly sourced from the respective original works, and the results with double daggers\textsuperscript{\ddag} are from~\citet{huang2023enhancing}.
    The results for \textsc{BrainStorm}, {\textsc{Algo}}, {\textsc{MBR-Exec}}, \textsc{CodeT}, and MPSC are based on GPT-3.5-Turbo. 
    The empty results for \textsc{CodeRanker}, \textsc{BrainStorm}, and {\textsc{Algo}} are due to reproducibility issues, as both the checkpoint and the full training data for each method are publicly unavailable.}
    \label{tab:humaneval}
\end{table*}

\paragraph{Baselines}
Throughout the benchmarks, we consider 
three baselines: GPT-3.5-Turbo and its
CoT prompting applied version of Self-planning~\cite{jiang2023selfplanning},
and \textsc{CodeT}~\cite{chen2023codet}.
For both HumanEval and CodeContests, we further use {three} code filtering methods---\textsc{CodeRanker}~\cite{inala2022faultaware}, {\textsc{MBR-Exec}~\cite{shi-etal-2022-learning}}, and MPSC~\cite{huang2023enhancing}---along 
with \textit{WizardCoder} 34B~\cite{luo2023wizardcoder} 
and GPT-4~\cite{openai2023gpt4}. 
For CodeContests, we additionally compare with {two CoT methods}: \textsc{Brainstorm}~\cite{li2023think} {and \textsc{Algo}~\cite{zhang2023algo}}.

\subsection{{\textbf{HumanEval-NFR: Embracing NFR Evaluation}}}
We introduce HumanEval-NFR benchmark, which is specifically designed to assess NFR satisfaction. 
It is an extension of HumanEval that additionally covers four NFR categories, chosen for their suitability for evaluation, through code execution using annotated test cases or automated assessment using existing metrics. Details on the annotation process and metrics we used are provided in Appendix~\ref{appendix:humaneval_nfr}.

Table~\ref{tab:humaneval_nfr} presents that \ours outperforms all baseline methods across various NFR categories except for maintainability.
Our conjecture is that, as which NFR categories to prioritize is uninformed in this experiment, \ours's consideration of all NFRs could potentially impede maintainability due to the influence of other categories.
We study the informed case of optimizing specific categories in Section~\ref{weighting}.
Across all approaches, satisfying the robustness category appears to be more difficult compared to other NFR categories, for which we provide further discussion in Appendix~\ref{appendix:robustness_performance}.

Notably, \ours~is desirable for evaluating all NFRs at once (i.e. All), 
outperforming \textsc{CodeT} with 22.16\% of Pass@$1$.

\subsection{{\textbf{HumanEval and CodeContests: Public Benchmarks for FR Evaluation}}}
We additionally report results on two popular code generation benchmarks targetting functional correctness.
HumanEval~\cite{codex} is a hand-crafted test benchmark with 164 programming problems along with public and hidden test cases.
CodeContests~\cite{li2022competition} consists of 13k/113/165 instances of train/valid/test data collected from multiple code competition websites.
While HumanEval tests the model's capability to implement rather simpler functions without errors, the competitive programming oriented nature of CodeContests often requires more complex form of reasoning such as algorithmic reasoning.
Each of these addresses different aspect of industrial software: 
the former is related to solving each of the simpler tasks composing larger and complex projects while the latter focuses on the logical and algorithmic perspective of software development.

In Table~\ref{tab:humaneval}, \ours~consistently outperforms the baseline methods. 
Specifically, on both benchmarks, \ours~leveraging GPT-3.5-Turbo, exceeds GPT-4's performance by a substantial margin of 4.81\%p and 10.45\%p in terms of Pass@$1$. 
In comparison with \textit{WizardCoder} 34B---a baseline that partially incorporates NFR considerations during the finetuning phase---\ours, which covers NFRs more comprehensively, achieves significantly higher performance. %
In CodeContests, while our custom GPT-3.5-Turbo + CoT prompting baseline is outdone by the state-of-the-art CoT methods \textsc{Brainstorm} {and \textsc{Algo}}, the application of \ours~outperforms {both approaches}, setting new state-of-the-art of Pass@$1$. 
We also compare \ours~with MPSC, a very recent baseline. Notably, \ours~surpasses MPSC in all Pass@$k$ metrics on HumanEval and Pass@$1$ on CodeContests, while \ours is much more cost-efficient.
We provide further discussion on computational costs in Section~\ref{analysis:testcase}.

\section{Analysis and Discussion}

\subsection{Efficiency and Effectiveness of Requirement-aware Test Case Generation}
\label{analysis:testcase}

\begin{figure}[t!]
    \centering  
    \includegraphics[width=0.9\linewidth]{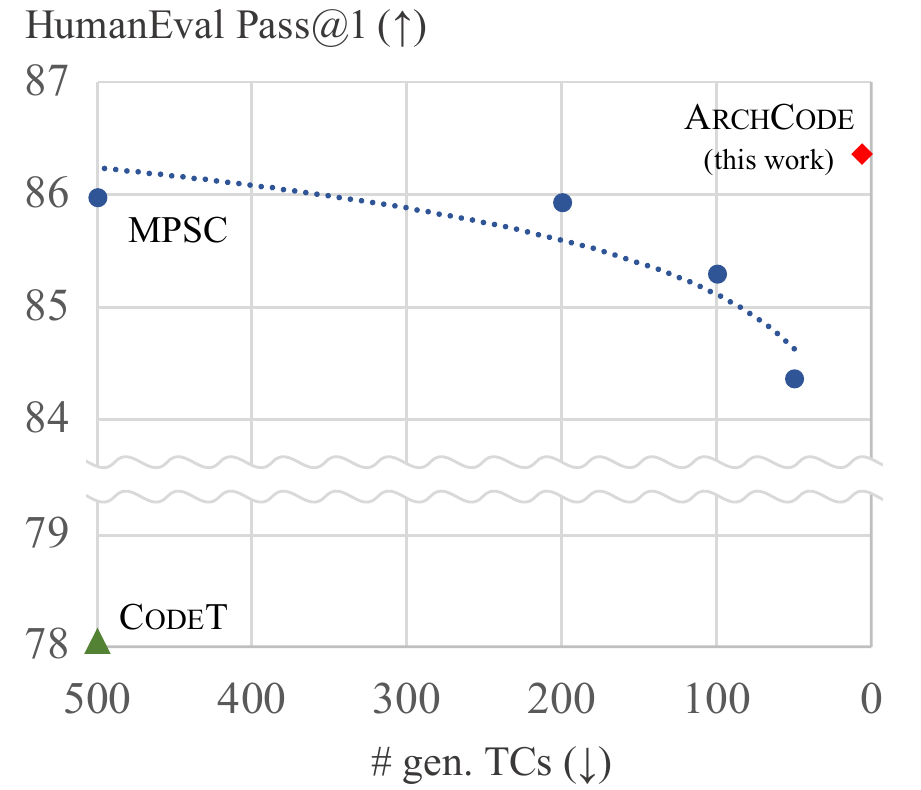}
    \caption{Pass@1 versus average number of test cases needed per problem on HumanEval. \ours~({\color{red}$\blackdiamond$}) achieves the highest Pass@1 score with significantly less number of generated test cases. All values are obtained from GPT-3.5-Turbo. The values for MPSC ({\color{blue}$\bullet$}) and \textsc{CodeT} ({\selectcolormodel{cmy}\color{green}$\blacktriangle$}) are from~\citet{huang2023enhancing}. {Best viewed in color.}} 
    \label{fig:n_test_cases}
\end{figure}

\paragraph{Efficiency}
In code filtering, a crucial step involves minimizing the number of generated test cases to reduce computational and time costs for code execution.
As shown in Figure~\ref{fig:n_test_cases}, existing approaches such as MPSC and \textsc{CodeT} requires to generate hundreds of test cases for performance. 
In contrast, \ours~targeting diverse requirement categories shows the best performance while {\textbf{significantly improving the efficiency by generating 50x smaller number of test cases. }}

\begin{table}[t]
    \centering
    \begin{tabular}{lccccccccc}
        \thickhline
         \multirow{2.6}{*}{Test Case} & 
         \multicolumn{3}{c}{\multirow{1.25}{*}{Code Gen. Method}} \\
           & 
         \multicolumn{3}{c}{\multirow{1.25}{*}{\ours}} \\
        \cmidrule(r){2-4}
         \rowcolor{white} \multirow{-2.4}{*}{Generation Method} & 
         \multirow{-1.25}{*}{$k$=1} & \multirow{-1.25}{*}{2} & \multirow{-1.25}{*}{5} \\\hline
        \textit{None} & 
            \phantom{0}{6.73} & 
            \phantom{0}9.79 & 
            14.63
             \\
        \textsc{CodeT} & 
            11.09 & 
            13.59 & 
            \underline{17.18}
             \\
        \textsc{CodeT}\small{ (w/o clustering)} &
        \underline{13.16} & \underline{14.14} & 16.48 \\
        \rowcolor{gray!10}
        \ours   & 
            \textbf{16.52} &
            \textbf{16.67} &
            \textbf{17.37}
            \\
        \thickhline
    \end{tabular}
    \caption{Code filtering results with different test case generation methods on CodeContests, while the code generation method is fixed to \ours.
    GPT-3.5-Turbo is used as the backbone model.
    MPSC is omitted as it is publicly unavailable.
    } %
    \label{tab:testcase_codecontests}
\end{table}

\newcolumntype{g}{>{\columncolor{gray!10}}c}
\begin{table*}[t]
    \centering
    \begin{tabular}{lccccccccc}
        \thickhline
         \multirow{2.6}{*}{Test Case} & 
         \multicolumn{9}{c}{\multirow{1.25}{*}{Code Generation Method}} \\
           & 
         \multicolumn{3}{c}{\multirow{1.25}{*}{GPT-3.5-Turbo}} & 
         \multicolumn{3}{c}{\multirow{1.25}{*}{GPT-3.5-Turbo + CoT}} & 
         \multicolumn{3}{c}{\multirow{1.25}{*}{\ours}} \\
        \cmidrule(r){2-4} \cmidrule(r){5-7} \cmidrule(r){8-10}
         \rowcolor{white} \multirow{-2.4}{*}{Generation Method} & 
         \multirow{-1.25}{*}{$k$=1} & \multirow{-1.25}{*}{2} & \multirow{-1.25}{*}{5} & 
         \multirow{-1.25}{*}{$k$=1} & \multirow{-1.25}{*}{2} & \multirow{-1.25}{*}{5} & 
         \multirow{-1.25}{*}{$k$=1} & \multirow{-1.25}{*}{2} & \multirow{-1.25}{*}{5} 
         \\\hline
        \textit{None} & 
            \phantom{0}2.62 & 
            \phantom{0}4.91 & 
            10.03 &
            \phantom{0}5.00 &  
            \phantom{0}7.90 & 
            \underline{12.08} &
            15.85 &
            20.23 &
            24.83
            \\
        \textsc{CodeT} & 
            \phantom{0}\underline{3.03} & 
            \phantom{0}4.83 & 
            10.03 &
            \phantom{0}\underline{5.50} &  
            \phantom{0}\underline{8.05} & 
            12.04 &
            \underline{17.00} &
            19.49 &
            24.69
            \\
        \textsc{CodeT}\small{ (w/o clustering)} & 
            \phantom{0}2.86 & 
            \phantom{0}\underline{5.02} & 
            \underline{10.30} &
            \phantom{0}4.69 &  
            \phantom{0}7.21 & 
            11.72 &
            16.09 &
            \underline{20.24} &
            \underline{25.07}
            \\
        \rowcolor{gray!10}
        \ours              & 
            \textbf{13.87} & 
            \textbf{14.53} & 
            \textbf{14.63} & 
            \textbf{13.82} & 
            \textbf{14.46} & 
            \textbf{14.52} & 
            \textbf{25.81} &
            \textbf{26.84} &
            \textbf{27.20} 
            \\
        \thickhline
    \end{tabular}
    \caption{Code filtering results with different test case generation methods on HumanEval-NFR (All).
    GPT-3.5-Turbo is used as the backbone model.
    MPSC is omitted as it is publicly unavailable.} %
    \label{tab:testcase_humaneval_nfr}
\end{table*}

\begin{table}[h]
    \centering
    \begin{tabular}{lccccccccc}
        \thickhline
         \multirow{2.6}{*}{Test Case} & 
         \multicolumn{3}{c}{\multirow{1.25}{*}{Code Gen. Method}} \\
           & 
         \multicolumn{3}{c}{\multirow{1.25}{*}{\ours}} \\
        \cmidrule(r){2-4}
         \rowcolor{white} \multirow{-2.4}{*}{Generation Method} & 
         \multirow{-1.25}{*}{$k$=1} & \multirow{-1.25}{*}{2} & \multirow{-1.25}{*}{5} 
         \\\hline
        \textit{None} & 
            75.06 &
            81.83 &
            87.95
            \\
        \textsc{CodeT} & 
            79.92 &
            87.63 &
            \textbf{91.00}
            \\
        \textsc{CodeT} {\small (w/o clustering)} & 
            \textbf{86.40} &
            \underline{88.21} &
            \underline{90.66}
            \\
        \rowcolor{gray!10}
        \ours              & 
            \underline{86.36} &
            \textbf{88.62} &
            90.48 
            \\
        \thickhline
    \end{tabular}
    \caption{Code filtering results with different test case generation methods  on HumanEval, while the code generation method is fixed to \ours.
    GPT-3.5-Turbo is used as the backbone model.
    MPSC is omitted as it is publicly unavailable. } %
    \label{tab:testcase_humaneval}
\end{table}

\paragraph{Effectiveness}
Tables~\ref{tab:testcase_codecontests},~\ref{tab:testcase_humaneval_nfr}, and~\ref{tab:testcase_humaneval} compare two test case generation methods, the naive way (\textsc{CodeT}) and \ours.
With the same code generation and filtering strategy applied, the latter generally outperforms the former with large margins, demonstrating the effectiveness of leveraging generated requirements to optimize test case generation. 
Meanwhile, \ours~yielded comparable results to \textsc{CodeT} without the use of clustering on HumanEval. 
We conjecture that for simpler benchmarks like HumanEval, \textsc{CodeT}'s approach of generating `general' test cases suffices.
While \textsc{CodeT} focuses on general test cases which are likely to have limited coverage, \textbf{\ours~distinctly promotes a diverse set of test cases targeting various requirement (sub)types.}

\subsection{Conditioning Code Generation on Test Cases}
\label{TDD}
In contrast to our approach of generating code and test cases in parallel and then applying subsequent postprocess mechanisms such as filtering, one can also consider conditioning the code generation on test cases,\footnote{One may also consider the opposite direction of generating the test cases conditioned on the generated code, which we do not visit in this paper.} taking inspiration from the Test-Driven Development (TDD;~\citealp{beck2022test}) methodology.
Table~\ref{tab:tdd} shows results consistent with those reported in \citet{chen2023codet}, indicating marginal improvement in performance is observed when conditioning code generation on the ground-truth and generated test cases, while
\textbf{incorporating software requirements through \ours~effectively boosts the score,} even without code filtering.
This suggests the overhead from introducing new sequential dependency in the generation process might not be worth the additional costs incurred.

\begin{table}[t]
    \centering
     \resizebox{0.48\textwidth}{!}{
    \begin{tabular}{lccc}
        \thickhline
           & 
         \multicolumn{3}{c}{\multirow{1.25}{*}{Pass@$k$}} \\
        \cmidrule(r){2-4}
         \rowcolor{white} \multirow{-2.4}{*}{Method} & 
         \multirow{-1.25}{*}{$k$=1} & \multirow{-1.25}{*}{2} & \multirow{-1.25}{*}{5} \\\hline        
         {GPT-3.5-Turbo} & {73.17} & {80.79} & {86.99} \\
        {+ Gold Test Cases} & {73.60} & {\underline{80.93}} & {\underline{87.24}} \\
        {+ Generated Test Cases} & {\underline{73.66}} & {80.54} & {86.53} \\
        \rowcolor{gray!10}
        {\ours $-$ Filtering} & {\textbf{75.06}} & {\textbf{81.83}} & {\textbf{87.95}} \\
        \thickhline
    \end{tabular}}
    \caption{Results on HumanEval with generating code conditioned additionally on test cases. While incurring sequential dependency and increased latency, TDD-like conditioning brings marginal improvement, as opposed to our method being effective even without filtering.}
    \label{tab:tdd}
\end{table}

\begin{figure}[t!]
    \centering  
    \includegraphics[width=0.9\linewidth]{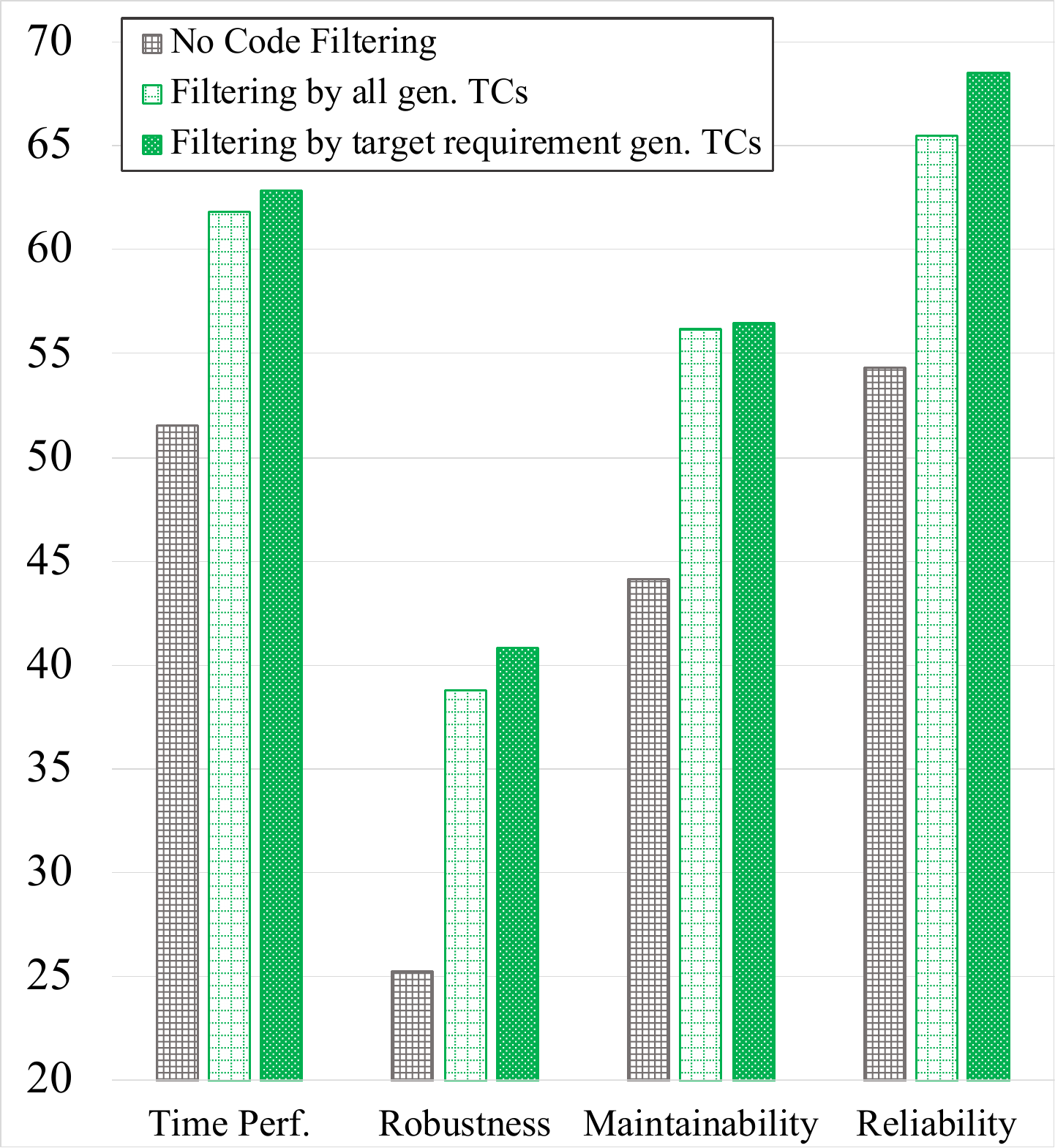}
    \caption{{Pass@$1$ score of \ours for each requirement category in HumanEval-NFR.
    Using dedicated test cases for filtering consistently outperforms blindly using all test cases. Best viewed in color.} }
    \label{fig:adaptive_weighting}
\end{figure}

\subsection{Preference over Requirements}
\label{weighting}
As mentioned before, \ours can be informed of any user preferences over the software requirements at code filtering time---after several code candidates have been generated and awaiting to be ranked.
Figure~\ref{fig:adaptive_weighting} presents the Pass@$1$ scores for each NFR category in the HumanEval-NFR benchmark, with different code filtering strategies applied. 
Using targeted test cases for reranking yields higher pass rates for that specific software requirement than using all test cases does.

Another approach to incorporate user preference over the requirements is to consider a subset of requirements when generating code or test cases, or to put emphasis on a subset of requirements by editing the prompt while presenting all requirements to the model.
We present detailed analyses for such scenarios in Appendix~\ref{appendix:preference_control}.

\subsection{\ours under Diverse Settings}
\label{generalizability}
Here we provide empirical results suggesting that \ours generalizes well to other models, datasets, etc. than those considered in the main experiments.

\paragraph{Open-source LLMs}
First, we showcase \ours combined with a relatively smaller model, namely \textit{WizardCoder} 7B. 
Table~\ref{tab:wizardcoder} indicates that applying \ours with the said backbone model leads to a notable 15.67\%p improvement in Pass@$1$ on HumanEval,
while incorporating in-context learning directly into \textit{WizardCoder} 7B itself has negative impacts.
Note that this observation is consistent with prior findings such as that in \citet{yuan2023evaluating}, that instruction tuning might compromise in-context learning capabilities of LLMs; \textit{WizardCoder} 7B is an instruction-tuned model based on \textsc{CodeLlama} 7B.

Meanwhile, in practical settings, diverse LLMs offer complementary benefits in terms of cost-performance trade-off, and thus mixing two models has been a conventional aproach to explore cost-performance design space~\cite{sun-etal-2023-chatgpt,wang2023voyager}. 
\textsc{ArchCode}${}_{\text{MIX}}$ shown in Tables~\ref{tab:wizardcoder} and~\ref{tab:multiple_java} similarly capitalizes on this space by directing most of the generation calls to affordable LLMs, while selectively delegating the part requiring the most of the reasoning capabilities to stronger ones.

\begin{table}[t]
    \centering
    \begin{tabular}{lccccccccc}
        \thickhline
        \multirow{2.25}{*}{Method} &  \multicolumn{3}{c}{\multirow{1.25}{*}{Pass@$k$}} \\
        \cmidrule(r){2-4}
          & 
         \multirow{-1.25}{*}{$k$=1} & \multirow{-1.25}{*}{2} & \multirow{-1.25}{*}{5} \\\hline
         \multicolumn{4}{c}{w/o ICL} \\\hdashline
        \textit{WizardCoder} 7B & 
            48.54 & 
            60.08 & 
            72.48 \\\hline
         \multicolumn{4}{c}{w/ ICL} \\\hdashline

         \textit{WizardCoder} 7B & 43.35 & 53.74 & 65.42 \\
         \phantom{ }+ CoT & 42.01 & 53.74 & 67.27 \\
         \rowcolor{gray!10}
         {\ours} & {\underline{64.21}} & {\underline{67.72}} & {\underline{72.84}} \\
         
         \rowcolor{gray!10}
        \textsc{ArchCode}${}_{\textrm{MIX}}$   & 
            \textbf{68.33} &
            \textbf{71.36} &
            \textbf{74.44}
            \\
        \thickhline
    \end{tabular}
    \caption{Experimental results using \textit{WizardCoder} 7B for code generation on HumanEval. `w/ ICL' means that 1-shot in-context learning is employed. `\textsc{ArchCode}${}_{\textrm{MIX}}$' indicates that code filtering is applied with test cases generated by \ours~using GPT-3.5-Turbo. 
    }
    \label{tab:wizardcoder}
\end{table}

\begin{table}[t]
    \centering
    \begin{tabular}{lc}
        \thickhline
         Method & Pass@$1$  \\\hline
        SantaCoder 1B & 15.00 \\
        +\phantom{ }SFT & \underline{18.16} \\
        \rowcolor{gray!10}
        \textsc{ArchCode}${}_{\textrm{MIX}}$ & \textbf{24.61}\\
        \thickhline
    \end{tabular}
    \caption{Experimental results on MultiPL-E java. `SFT' denotes sparse fine-tuning. `\textsc{ArchCode}${}_{\textrm{MIX}}$' indicates that code filtering is applied with test cases generated by \ours~using GPT-3.5-Turbo.}
    \label{tab:multiple_java}
\end{table}

\paragraph{Other Programming Languages}
We also extend the evaluation of \ours to the task of Java code generation, using the MultiPL-E~\cite{cassano2022multipl} benchmark and the backbone model SantaCoder 1B~\cite{allal2023santacoder}.
To address the rather limited capacity of a smaller model, we further applied sparse fine-tuning (\citet{ansell-etal-2022-composable}; SFT) on a public Java train set.
We provide more details in Appendix~\ref{appendix:generalizability}.
The results in Table~\ref{tab:multiple_java} demonstrate the effectiveness of the proposed method in generating Java code, supporting that our method is generally applicable to programming languages other than Python.

\section{Conclusion}
We proposed \ours, a framework incorporating software requirements from textual descriptions for LLM-based code generation. 
This systematic approach not only identifies these requirements but also harnesses them to guide the code generation process. The verification of code snippets with the generated test cases tailored to each requirement provides a robust validation layer for the alignment with detected requirements.
On HumanEval and CodeContests, \ours~with GPT-3.5-Turbo exceeded GPT-4's performance by 4.81\%p and 10.45\%p of Pass@$1$. 
\ours~requires 50x less generated test cases compared to MPSC and \textsc{CodeT}, while outperforming them.
In addition, we introduced a new benchmark named HumanEval-NFR for evaluating how well LLMs can pursue non-functional requirements in code generation task.
Further analysis shows the pertinence of parallel generation of code and test case, and the efficiency and the effectiveness of \ours's requirement-aware test case generation. 

\section*{Acknowledgment}
This work was supported by LG AI Research, and partly supported by Electronics and Telecommunications Research Institute (ETRI) grant funded by ICT R\&D program of MSIT/IITP (2022-0-00995, Automated reliable source code generation from natural language descriptions).

\section*{{\textbf{Limitations}}}

{\ours~leverages in-context learning as a tool to integrate both functional and non-functional requirements in the processes of code and test case generation. We did not studied prompt engineering and devising more sophisticated in-context examples which is beyond the scope of this work.}
 
{\ours~encompassed three functional and four non-functional requirements, aligning with the established taxonomy within software engineering literature~\cite{glinz2007non}. 
However, the potential for future work lies in addressing more complex and varied requirements involving larger pieces of code, as well as accommodating changes in software requirements over time.}

{Lastly, as \ours relies on generated requirements to guide subsequent code and test case generation process, although qualitative analysis suggests its impact could be limited in practice, additional measures to mitigate cascading errors via human intervention or self-correction by LLMs, etc.~\cite{shinn2023reflexion,wang2023voyager,yao2023tree,chen2024teaching} can be necessitated.}

\section*{{\textbf{Ethical and Social Implications}}}
{\ours~leverages LLMs to automatically generate software requirements, code, and test cases, thereby enhancing productivity and reducing manual labor for developers. 
However, to maximize these advantages while addressing potential risks, such as the creation of code with safety or security vulnerabilities as discussed in~\citet{codex}, careful consideration is essential. 
Strategies to mitigate these risks include establishing default requirements for desired outcomes, delineating the permissible scope of generated code, and ensuring that the code remains within its authorized boundaries. }

\bibliography{anthology,custom}
\bibliographystyle{acl_natbib}

\appendix
\clearpage

\section{Implementation Details}
\label{impl_detail}
We used \texttt{gpt-3.5-turbo-16k}~\cite{openai2022chatgpt} as the backbone LLM for most of the experiments, with ICL and nucleus sampling~\cite{Holtzman2020The} with $p=0.95$ and temperature $T=0.8$ following~\citet{codex,nijkamp2023codegen,chen2023codet}.
We used different in-context examples for each benchmark: a single HumanEval-style (problem description, code) pair from~\citet{li2023think} for HumanEval-NFR, eight pairs from the training set of the MBPP~\cite{DBLP:journals/corr/abs-2108-07732} benchmark for HumanEval~\cite{codex}. For CodeContests~\cite{li2022competition} we used a single pair from the train set.

To apply CoT prompting~\cite{kojima2022large,10.5555/3618408.3619681,zhang2023automatic}, as the state-of-the-art methods \textsc{Brainstorm}~\cite{li2023think} and \textsc{Algo}~\cite{zhang2023algo} are publicly unavailable, we generated the reasoning chains of code outline by using self-planning~\cite{jiang2023selfplanning}. However, directly using the reasoning chains provided by Self-planning can result in data contamination on HumanEval because these chains are based on the test examples. Thus, rather using them directly, we utilized them to generate the reasoning chains for the aforementioned in-context examples, then used the generated reasoning chains for ICL. 

\ours~uses three reasoning chains when generating code: the initial program outline (the same reasoning chains as in GPT-3.5-Turbo + CoT), requirements described in Subsection~\ref{subsec:software_requirements} and Appendix~\ref{appendix:generated_requirements}, and the final program outline---the revised version of the initial program outline, modified to meet the requirements.

We generated $n=10$ code samples for every problem in the benchmarks.
To enhance the diversity of the generated code, we employed nucleus sampling to produce $n$ initial program outlines induced from Self-planning.
The rest of the reasoning chains were concurrently generated using greedy sampling, culminating in a total of $n$ final code outputs.

Our implementation is largely based on the LangChain library.\footnote{\url{https://github.com/langchain-ai/langchain}} 
Regarding the execution and evaluation of the generated code,
we modified some code from the CodeEval repository\footnote{\url{https://huggingface.co/spaces/evaluate-metric/code_eval}} which is available on Huggingface.

Regarding the alignment of sub-requirements to the corresponding test cases for code filtering, we generate all test cases for all sub-requirements in one iteration to minimize LLM calls, as shown in Tables~\ref{tab:gen_rq} and~\ref{tab:gen_tcs}. 
This approach leverages formatted in-context examples from Tables~\ref{tab:ice_rq_gen} and~\ref{tab:ice_tc_gen}. 
Subsequently, test cases are parsed and categorized according to corresponding sub-requirement types, followed by a test run with generated code snippets for code filtering.

\subsection{Open-sourced Backbone Models and Java Language}
\label{appendix:generalizability}
\begin{table}[t]
    \centering
    \begin{tabular}{lccccccccc}
        \thickhline
        \multirow{2.25}{*}{Method} &  \multicolumn{3}{c}{\multirow{1.25}{*}{Pass@$k$}} \\
        \cmidrule(r){2-4}
          & 
         \multirow{-1.25}{*}{$k$=1} & \multirow{-1.25}{*}{2} & \multirow{-1.25}{*}{5} \\\hline
        \textit{WizardCoder} 7B & 
            1.21 &
            2.05 &
            3.35\\
        \rowcolor{gray!10}
        \textsc{ArchCode}${}_{\textrm{MIX}}$  & 
            \textbf{4.24} & 
            \textbf{4.24} & 
            \textbf{4.24} 
            \\
        \thickhline
    \end{tabular}
    \caption{Experimental results using \textit{WizardCoder} 7B w/o ICL for code generation on CodeContests.
    `\textsc{ArchCode}${}_{\textrm{MIX}}$' indicates that code filtering is applied with test cases generated by \ours~using GPT-3.5-Turbo.
    } %
    \label{tab:wizardcoder_cc}
\end{table}

\paragraph{Open-source LLMs}
We utilized huggingface's text-generation-inference
\footnote{\url{https://huggingface.co/docs/text-generation-inference}} 
to parallelize \textit{WizardCoder} 7B on two NVIDIA RTX A6000 48GBs 
for inference purposes exclusively.
It took approximately one hour to experiment with one method on the entire HumanEval benchmark.
Consistent to the results on HumanEval shown in Table~\ref{tab:wizardcoder}, 
Table~\ref{tab:wizardcoder_cc} also shows that 
\ours~significantly contributes to Pass@$k$ scores on CodeContests.  

\paragraph{Other Programming Languages} For sparse fine-tuning, 
we followed~\citet{ansell-etal-2022-composable} to train 3\% of the SantaCoder 1B~\cite{allal2023santacoder}  parameters with the batch size of 8 (1*grad\_accum of 8), the learning rate of 2e-5, the L1 regularization of 0, and the max training epochs of 3, using a single NVIDIA RTX A6000 48GB for 2 hours. 
For the training set, we utilized MegaCodeTraining,\footnote{\url{https://huggingface.co/datasets/rombodawg/MegaCodeTraining}} a public dataset set, while using java related data only. 

\section{HumanEval-NFR Construction}
\label{appendix:humaneval_nfr}

\begin{table}[t]
    \centering
    \begin{tabular}{llc}
        \thickhline
        Requirements & Subtype & {\small \# GT TCs} \\\hline
        \multicolumn{3}{c}{from HumanEval}\\
        \hdashline
        Functional & General + Edge   & 8.1 \\\hline
        \multicolumn{3}{c}{Additionally Annotated}\\
        \hdashline
        \multirow{2}{*}{Functional} & General   & 3.1 \\
                                    & Edge      & 2.6 \\\hdashline
        \multirow{3}{*}{Non-Functional} & Time Perf. & 1.8 \\
                                    & Robustness & 2.3 \\
                                    & Maintainability & 1.0 \\
        \thickhline
    \end{tabular}
    \caption{The average number of ground truth test cases (GT TCs) per problem on HumanEval-NFR. Note that reliability is confirmed by checking whether other test cases completed gracefully (without errors), regardless of whether the output was correct. Regarding the maintainability, one test case was sufficient as we defined it as whether the generated code exhibits the specified level of Cyclomatic Complexity or not.} 
    \label{tab:n_test_cases_humaneval_nfr}
\end{table}

The HumanEval-NFR benchmark, an extension of HumanEval~\cite{codex}, evaluates both FRs and NFRs of code. 
HumanEval-NFR comprises the same 164 problems as in the original HumanEval suite.
While encompassing all the problem descriptions and ground truth test cases from the original HumanEval benchmark for FR verification, it introduces additional test cases for FR and NFR verification.
The statistics of HumanEval-NFR's ground truth test cases are shown in Table~\ref{tab:n_test_cases_humaneval_nfr}.

Writing new ground truth test cases involved a two-step process. 
First, we generated candidate test cases based on the existing HumanEval problems using \ours based on GPT-3.5-Turbo. 
Second, we revised those test cases both in automatic or manual manner to ensure the quality of the test suite, based on the following protocols.

\subsection{Quality Control for FR Test Cases}
For candidate test cases evaluating FRs, we executed the ground truth code from the original HumanEval benchmark against each test case tailored to functional requirements. 
Those that the ground truth code does not pass were discarded.

\subsection{Quality Control for NFR Test Cases}
For candidate test cases verifying NFRs, three authors manually validated the quality of generated test cases. 
During validation, the authors adhered to the following principles:
\begin{itemize}
    \item Misclassified test cases should be rectified, and any duplicates should be eliminated.
    \item Test cases should compatible to the original ground truth code.
    If any discrepancy is found in the code, or if a test case is deemed impractical or overly complex, adjustments should be made to ensure it aligns with the original problem description.
\end{itemize}

\noindent In addition, the authors consider guidelines specific to each NFR category:
\paragraph{Time Performance} 
As \textit{Rice's Theorem}~\cite{rice1953classes} states, all non-trivial properties of Turing-recognizable languages are undecidable, which in essence means that there could be no `time-complexity checkers.'
Therefore, HumanEval-NFR follows conventional strategies used in competitive programming contests, such as Codeforces,\footnote{\url{https://codeforces.com}} where code is executed with relatively larger inputs to ensure that inefficient implementations cannot complete within the specified timeout. 
Specifically, we set the timeout as 5 seconds for all problems.

\paragraph{Robustness} 
Test cases for this category verify whether the implementation gracefully handles diverse types of invalid inputs, such as a string passed as an argument where an integer is expected. 
For technical reasons, we expect the code to return values like \texttt{None}, an empty list, or \texttt{False}---all of which are logically evaluated as False in the Python language---rather than forcing it to raise exceptions or using any other means to indicate it has detected an abnormal input.

\paragraph{Maintainability} 
{To validate maintainability, we consider code complexity, which affects the ease of understanding and updating the code~\cite{magel1982applying}. }
Specifically, HumanEval-NFR computes the Cyclomatic Complexity (CC;~\citealp{mccabe1976complexity}) of code, which evaluates code {complexity} by accounting for the depth of nested indented blocks, then checks whether the observed CC score is lower than the threshold. 
The threshold is set to 5 if the ground truth code from the original HumanEval benchmark has a CC value below 5; if the CC value exceeds 5, we set the threshold as 10~\cite{watson1996structured}.

\paragraph{Reliability} 
Rather than generating dedicated test cases, HumanEval-NFR assesses code reliability by executing all the ground truth test cases for the problem and checks if any runtime errors are raised, without verifying if the outputs are correct. 
This approach aligns with the category's focus on minimizing system failures and extending the mean-time-to-failure.

\begin{table*}[htb!]
    \centering
    \begin{tabular}{lccccccccccccccc}
        \thickhline
         & 
         \multicolumn{2}{c}{\multirow{1.25}{*}{All}} & 
         \multicolumn{2}{c}{\multirow{1.25}{*}{Time Perf.}} &
         \multicolumn{2}{c}{\multirow{1.25}{*}{Robustness}} &
         \multicolumn{2}{c}{\multirow{1.25}{*}{Maintainability}} &
         \multicolumn{2}{c}{\multirow{1.25}{*}{Reliability}} 
         \\
        \cmidrule(r){2-3}\cmidrule(r){4-5}\cmidrule(r){6-7}\cmidrule(r){8-9}\cmidrule(r){10-11}
         \multirow{-1.25}{*}{Pass@$k$} 
         & \multirow{-1.25}{*}{$k$=1} & \multirow{-1.25}{*}{5} &
          \multirow{-1.25}{*}{$k$=1} & \multirow{-1.25}{*}{5} &
          \multirow{-1.25}{*}{$k$=1} & \multirow{-1.25}{*}{5} &
          \multirow{-1.25}{*}{$k$=1} & \multirow{-1.25}{*}{5} &
          \multirow{-1.25}{*}{$k$=1} & \multirow{-1.25}{*}{5} 
          \\\hline
        GPT-3.5-Turbo & 
            {\small \phantom{0}{2.62}} & 
            {\small 10.03 } & {\small 53.48} & {\small 65.75} & {\small \phantom{0}4.21} & {\small 14.55} & {\small 53.23} & {\small \textbf{68.38}} & {\small 20.98} & {\small 36.72}
             \\
        \phantom{ }\phantom{ }+ NFR Instruction & 
            {\small 10.70} &
            {\small 26.30} & {\small 48.23} & {\small 63.22} & {\small 14.02} & {\small 34.65} & {\small 43.54} & {\small 60.73} & {\small 62.07} & {\small 90.79}
            \\\hdashline
        GPT-3.5-Turbo + CoT &
            {\small \phantom{0}5.00} &
            {\small 12.08} & {\small 50.00} & {\small 66.03} & {\small \phantom{0}7.32} & {\small 17.67} & {\small 44.33} & {\small 62.00} & {\small 45.49} & {\small 66.83}
            \\
        \phantom{ }\phantom{ }+ NFR Instruction & 
            {\small \phantom{0}5.30} &
            {\small 17.50} & {\small 50.00} & {\small 66.03} & {\small \phantom{0}7.62} & {\small 24.04} & {\small 43.66} & {\small 61.87} & {\small 65.55} & {\small 93.64}
            \\\hdashline
        \rowcolor{gray!10}
        \ours $-$ Filtering   & 
            {\small 15.85} &
            {\small 24.83} & {\small 51.52} & {\small 65.87} & {\small 25.24} & {\small 38.82} & {\small 44.15} & {\small 59.41} & {\small 54.33} & {\small 70.39}
            \\
        \rowcolor{gray!10}
        \phantom{ }\phantom{ }+ NFR Instruction &
            {\small 19.80} &
            {\small \underline{30.20}} & {\small 51.34} & {\small 65.24} & {\small 28.66} & {\small \underline{43.37}} & {\small 44.33} & {\small 57.69} & {\small \underline{88.48}} & {\small \textbf{95.56}}
            \\\hdashline
        \rowcolor{gray!10}
        \ours   & 
            {\small \underline{25.19}} &
                {\small 27.33 } & {\small \underline{62.86}} & {\small \textbf{69.70}} & {\small \underline{40.86}} & {\small 42.72} & {\small \textbf{56.43}} & {\small \underline{62.23}} & {\small 68.53} & {\small 74.67}
            \\
        \rowcolor{gray!10}
        \phantom{ }\phantom{ }+ NFR Instruction &
            {\small\textbf{29.50}} &
            {\small \textbf{32.88}} & {\small \textbf{62.99}} & {\small \underline{67.10}} & {\small \textbf{43.46}} & {\small \textbf{47.13}} & {\small \underline{54.21}} & {\small 59.42} & {\small \textbf{92.46}} & {\small \underline{95.01}}\\
        \thickhline
    \end{tabular}
    \caption{Experimental results of requirements instruction prompting on HumanEval-NFR. Boldfaced and underlined values indicate the 1st and 2nd largest scores, respectively. `+ NFR Instruction' means that the further prompt engineered instruction for NFR consideration shown in Table~\ref{tab:nfr_instruction} is applied. `$-$ Filtering' denotes an ablated version of \ours, without code filtering.}
    \label{tab:nfr_instruction_ablation}
\end{table*}

\begin{table}[h]
\begin{tabularx}{0.48\textwidth}{X}
\hline
\# Code must satisfy not only functional requirements but also the following non-functional requirements.\\
\# Non-functional Requirements\\
\#\# Performance: Pertains to time-centric aspects such as algorithmic time complexity or stipulated timeout conditions.\\
\#\# Robustness: Ensures that code is resilient to invalid inputs.\\
\#\# Maintainability: Considers factors that contribute to the ease of maintenance.\\
\#\# Reliability: Ensures that code can handle errors gracefully without causing system failures over an extended period.\\
Write a code for the problem.\\
\hline
\end{tabularx}
\caption{Engineered prompt which further specified the details of each NFR, used in Table~\ref{tab:nfr_instruction_ablation}.} 
\label{tab:nfr_instruction}
\end{table}

\section{Gains from Prompt Engineering}
\label{nfr_instruction_ablation}
In this study, we did not focus on devising sophisticated prompts, as our main contribution does not rely heavily on using prompt-engineered instructions.
Therefore, we can expect even more performance gains when the prompt is further engineered as in Table~\ref{tab:nfr_instruction}, as we intentionally kept prompt simple in our main experiments. 

Table~\ref{tab:nfr_instruction_ablation} shows that \ours~is scalable to requirement instruction prompts, showing the best performance on HumanEval-NFR (All) when both are applied. 
Unlike CoT and NFR Instruction that improve Robustness and Reliability only, \ours~contributes to all NFR types. Notably, time performance and maintainability are enhanced solely by \ours's code filtering, highlighting the unique contribution over prompt engineering. 

\section{Correctness of Generated Requirements}
\label{appendix:generated_requirements}
\begin{figure}[h!]
    \centering  
    \hspace*{-0.3cm}
    \includegraphics[width=1.0\linewidth]{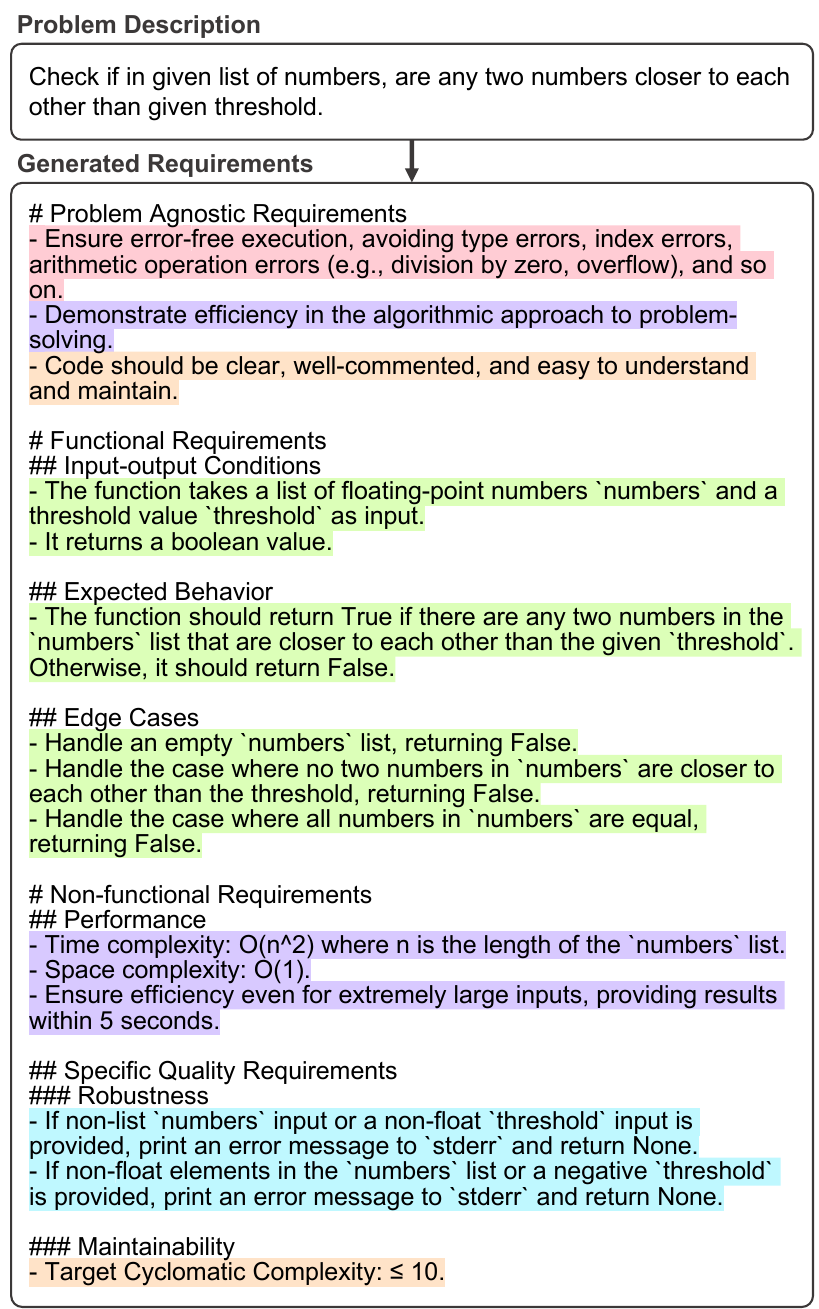}
    \caption{A real-life example of generated requirements from \texttt{HumanEval-NFR/0} by \ours. Best viewed in color.} 
    \label{fig:requirements}
\end{figure}

\paragraph{Format} 
As presented in Figure~\ref{fig:requirements}, we organized the structure of software requirements into two parts: problem-agnostic and problem-specific.
The former describes general guidelines throughout problems related to reliability, performance, and maintainability. 
The latter includes more specific instructions depending on the problem description, including all three subtypes of functional requirements, the target time complexity for time performance, the invalid conditions for robustness, and the target Cyclomatic Complexity for maintainability. 

\paragraph{Validation}
\label{appendix:errors_in_requirements}
To confirm the correctness of the requirements generated by \ours with GPT-3.5-Turbo, we randomly selected three problems, one for each of the following categories: (1) all,  (2) some or (3) none of the generated code samples passed the tests.
We manually verified validity of each set of requirements for each case, of which the results are summarized in Table~\ref{tab:gen_reqs_analysis}. 
Surprisingly, all the generated requirements were correct, regardless of the corresponding generated code's correctness.

\section{Analysis of Generated Test Cases by \ours}
\label{appendix:analysis_of_gen_tcs}
\begin{table}[h]
    \centering
    \begin{tabular}{llc}
        \thickhline
        Requirements & Subtype & {\small \# gen. TCs} \\\hline
        \multirow{2}{*}{Functional} & General   & 3.1 \\
                                    & Edge      & 3.1 \\\hdashline
        \multirow{3}{*}{Non-Functional} & Time Perf. & 1.9 \\
                                    & Robustness & 2.0 \\
                                    & Maintainability & 1.0 \\
        \thickhline
    \end{tabular}
    \caption{The average number of generated test cases by \ours~per problem on HumanEval-NFR. Note that reliability is confirmed by checking whether other test cases completed gracefully (without errors), regardless of whether the output was correct. Regarding the maintainability, one test case was sufficient as we defined it as whether the generated code exhibits the specified level of Cyclomatic Complexity or not.} 
    \label{tab:n_test_cases}
\end{table}

\begin{table}[h]
    \centering
    \begin{tabular}{llc}
        \thickhline
        Method  & Subtype & Acc. \\\hline
        \multicolumn{3}{c}{HumanEval} \\\hdashline
        \rowcolor{gray!10}                          & General & 89\% \\
        \rowcolor{gray!10}
        \multirow{-2}{*}{\ours}  & Edge    & 49\% \\\hline
        \multicolumn{3}{c}{CodeContests} \\\hdashline
        \rowcolor{gray!10}                          & General & 64\% \\
        \rowcolor{gray!10}
        \multirow{-2}{*}{\ours}  & Edge    & 18\% \\
        \thickhline
    \end{tabular}
    \caption{The pass ratio of the ground-truth code against generated test cases tailored to functional requirements on HumanEval. } 
    \label{tab:acc_test_cases}
\end{table}

Table~\ref{tab:n_test_cases} shows the average number of generated test cases by \ours for each requirement category. 
Table~\ref{tab:acc_test_cases} reports the accuracy of generated test cases tailored to functional requirements. 
Although the accuracy of generated edge cases is relatively low, they still play a key role in code filtering as evidentiated by the performance discrepancy between \ours~and \textsc{CodeT} (w/o clustering) presented in Tables~\ref{tab:testcase_codecontests}.
It is noteworthy that \textsc{CodeT} generates `general' test cases, and the results for \ours~and \textsc{CodeT} are comparable, given that both methods are based on the same GPT-3.5-Turbo architecture.
We conjecture that the relatively low accuracy of generated edge cases does not prevent them from being substantially useful in code filtering, for that wrongful test cases tend to accept or to reject both the correct and incorrect code, rather than selectively passing incorrect ones.
In other words, the overall ranking of the generated code samples is hardly affected by the wrongly generated test cases.

For the validation of non-functional requirements, we can implicitly confirm through Figure~\ref{fig:adaptive_weighting} as using targeted test cases (filled green) yielded better results than using all test cases (empty green) across all NFR categories, as mentioned in Section~\ref{weighting}.

\section{NFR Preference Control}
\label{appendix:preference_control}

Unlike for FRs, one can consider adopting preferences among NFRs, as (1) they inherently describe rather `optional' tweaks that can additionally guide the behavior of the code and (2) some trade-off relationships among different NFRs~\cite{chung2012non,krummenacher2007towards,moreira2005multi,gross2001non}. 
An example of the latter would be the trade-off between the time performance and the rest of the NFRs~\cite{krummenacher2007towards,gross2001non}.

We have already shown in Section~\ref{weighting} that such control can be achieved by adjusting the weights of test cases for different NFR categories in the adjusting approach.

\begin{table*}[h]
    \centering
    \resizebox{\textwidth}{!}{
    \begin{tabular}{llcccccccccc}
        \thickhline
         \multirow{1.5}{*}{\small NFR(s) shown in} & 
         \multirow{1.5}{*}{\small Preferred} & 
         \multicolumn{2}{c}{\multirow{1.25}{*}{\small All}} & 
         \multicolumn{2}{c}{\multirow{1.25}{*}{\small Time Perf.}} &
         \multicolumn{2}{c}{\multirow{1.25}{*}{\small Robustness}} &
         \multicolumn{2}{c}{\multirow{1.25}{*}{\small Maintainability}} &
         \multicolumn{2}{c}{\multirow{1.25}{*}{\small Reliability}} 
         \\
        \cmidrule(r){3-4}\cmidrule(r){5-6}\cmidrule(r){7-8}\cmidrule(r){9-10}\cmidrule(r){11-12}
         \multirow{-1.5}{*}{\small few-shot examples} &
         \multirow{-1.5}{*}{\small NFR(s)} 
         & \multirow{-1.25}{*}{\small $k$=1} & \multirow{-1.25}{*}{\small 5} &
          \multirow{-1.25}{*}{\small $k$=1} & \multirow{-1.25}{*}{\small 5} &
          \multirow{-1.25}{*}{\small $k$=1} & \multirow{-1.25}{*}{\small 5} &
          \multirow{-1.25}{*}{\small $k$=1} & \multirow{-1.25}{*}{\small 5} &
          \multirow{-1.25}{*}{\small $k$=1} & \multirow{-1.25}{*}{\small 5} 
          \\\hline
        \multicolumn{12}{c}{No Preference}\\\hdashline
        \textit{\small All} & \textit{\small None} & {\small \underline{15.85}} & {\small 24.83} & {\small 51.52} & {\small 65.87} & {\small 25.24} & {\small 38.82} & {\small 44.15} & {\small 59.41} & {\small 54.33} & {\small 70.39}
            \\\hline
        \multicolumn{12}{c}{Preference Control by Instruction}\\\hdashline
        \textit{\small All} & \textit{\small Time Perf.} & {\small 15.79} & {\small 26.50} & {\small \underline{52.38}} & {\small \textbf{68.96}} & {\small 25.30} & {\small 40.22} & {\small 44.15} & {\small \underline{63.40}} & {\small \textbf{86.10}} & {\small \textbf{95.18}}
            \\
        \textit{\small All} & \textit{\small All - Time Perf.} & {\small 15.55} & {\small \underline{26.75}} & {\small 51.28} & {\small 65.56} & {\small \underline{25.61}} & {\small \underline{41.02}} & {\small \underline{45.55}} & {\small 60.94} & {\small \underline{82.44}} & {\small \underline{94.13}}
            \\\hline
        \multicolumn{12}{c}{Preference Control by Plug-and-Play}\\\hdashline
        \textit{\small Time Perf.} & \textit{\small Time Perf.} & {\small \phantom{0}3.84} & {\small \phantom{0}7.85} & {\small \textbf{53.29}} & {\small \underline{67.55}} & {\small \phantom{0}7.93} & {\small 15.45} & {\small \textbf{49.45}} & {\small \textbf{66.65}} & {\small 35.98} & {\small 56.24} \\
        \textit{\small All - Time Perf.} & \textit{\small All - Time Perf.} & {\small \textbf{17.50}} & {\small \textbf{28.74}} & {\small 52.01} & {\small 66.69} & {\small \textbf{27.68}} & {\small \textbf{43.63}} & {\small 44.15} & {\small 60.22} & {\small 49.82} & {\small 68.13} \\
        \thickhline
    \end{tabular}}
    \caption{NFR preference control of \ours~without applying code filtering (i.e. \ours~$-$ Filtering). Boldfaced and underlined values indicate the 1st and 2nd largest scores, respectively. \textit{All} means all NFRs---time performance, robustness, maintainability, and reliability---are targeted, and \textit{All - Time Perf.} means all NFRs except for time performance are targeted.
    In the `Preference Control by Instruction' setting, all NFRs are included in the prompt, and an additional instruction to prioritize specific NFR(s) is appended. 
    In the `Plug-and-Play' setting, only targeted NFRs are included in the few-shot examples, while no preference among them is assumed.}
    \label{tab:nfr_preference_exploration}
\end{table*}

Here, we present an alternative means of accomodating preference, by guiding the generation of code and test cases in a preference-aware manner as with the following two methods:
\begin{itemize}
    \item \textbf{Preference Control by Instruction}: We include all NFRs in prompts just as before, while explicitly expressing the preference at the end (e.g. \textit{``Consider the time performance requirement to be the most important.''}). 
    \item \textbf{Preference Control by Plug-and-Play}: We only present the preferred NFRs in prompts, without an explicit description of the preference. All included NFRs are considered equally important, with no prioritization among them. 
\end{itemize}

Table~\ref{tab:nfr_preference_exploration} shows that the plug-and-play approach inflicts a larger impact on Pass@$k$ on HumanEval-NFR (All), compared to the instruction-based method.
Notably, the plug-and-play approach considering all but time performance showed the best Pass@$k$ scores, which we attribute to the trade-off between time performance and the rest of the NFRs: focusing on the other requirements is relatively free of negative interference that would hurt the performance.
In the categories of time performance and robustness, both instruction and plug-and-play preference settings showed improvements when each of them was targeted. 
For the maintainability category, the best results were observed when only time performance was preferred in the plug-and-play approach. 
This is likely due to the omission of code lines for handling exceptions related to invalid inputs, as the robustness category was not considered. 
In the case of reliability, which is assessed by the error-free execution of code across all test cases (without dedicated test cases), performance improvement was observed irrespective of preference in the instruction-based approach. 
As demonstrated in Table~\ref{tab:nfr_instruction_ablation}, this suggests that prompt engineering can reduce error ratios; we leave exploration towards this direction to future work.

\section{Varying Difficulty Levels of NFRs in HumanEval-NFR}
\label{appendix:robustness_performance}
Orthogonal to \ours's contribution towards satisfying every requirement category, we observe a general trend of relatively low performance in the robustness category compared to others in HumanEval-NFR, as shown in Table~\ref{tab:humaneval_nfr} and Figure~\ref{fig:adaptive_weighting}. 
One conjecture is that the difficulty lies among the NFRs as inferred from the original HumanEval benchmark.
As the ground truth code snippets are generally not complex, handling large input size (time performance), managing small Cyclomatic Complexity (maintainability), and avoiding runtime error while running other test cases (reliability) might be easier than correctly handling every possible invalid input (robustness).

\begin{table*}[h]
    \centering
    \begin{tabularx}{\textwidth}{llX}
        \thickhline
        \multirow{2}{*}{Index} & Generated Requirements  & Generated Code  \\
        & (manually validated) & (validated by Ground Truth Test Cases)\\\hline
        \multicolumn{3}{c}{\texttt{HumanEval-NFR/51}}\\\hdashline
        0 & Correct & Passed\\
        1 & Correct & Passed\\
        2 & Correct & Passed\\
        3 & Correct & Passed\\
        4 & Correct & Passed\\
        5 & Correct & Passed\\
        6 & Correct & Passed\\
        7 & Correct & Passed\\
        8 & Correct & Passed\\
        9 & Correct & Passed\\\hline
        \multicolumn{3}{c}{\texttt{HumanEval-NFR/11}}\\\hdashline 
        0 & Correct & failed: invalid literal for int() with base 2: `'\\
        1 & Correct & failed: Failed to handle input strings of different lengths.\\
        2 & Correct & failed: Failed to handle large input size.\\
        3 & Correct & failed: Failed to handle input strings of different lengths.\\
        4 & Correct & Passed\\
        5 & Correct & failed: Failed to have a Cyclomatic Complexity less than or equal to 10 by Radon.\\
        6 & Correct & failed: Failed to handle input strings of different lengths.\\
        7 & Correct & Passed\\
        8 & Correct & failed: invalid literal for int() with base 2: `'\\
        9 & Correct & Passed\\\hline
        \multicolumn{3}{c}{\texttt{HumanEval-NFR/0}}\\\hdashline 
        0 & Correct & failed: Failed to handle large input size.\\
        1 & Correct & failed: Failed to handle large input size.\\
        2 & Correct & failed: Failed to handle large input size.\\
        3 & Correct & failed: Failed to handle large input size.\\
        4 & Correct & failed: Failed to handle large input size.\\
        5 & Correct & failed: Failed to handle large input size.\\
        6 & Correct & failed: Failed to handle large input size.\\
        7 & Correct & failed: Failed to handle large input size.\\
        8 & Correct & failed: Failed to handle large input size.\\
        9 & Correct & failed: Failed to handle large input size.\\
        \thickhline
    \end{tabularx}
    \caption{Validation of generated requirements by \ours~on HumanEval-NFR. Note that each generated code snippet is uniquely tailored to correspond with a distinct set of requirements, ensuring that no code snippet shares the same set of requirements.} 
    \label{tab:gen_reqs_analysis}
\end{table*}

\begin{table*}[h]
    
    \centering
    \begin{tabularx}{\textwidth}{lXcX}
        \thickhline
        \multirow{2}{*}{\small{Sub-requirement}} & \multirow{2}{*}{\small{Generated Requirement}} & \small{Human} & \multirow{2}{*}{\small{Explanation}}  \\
        & & \small{Validation} & \\\hline
        \multicolumn{4}{c}{\texttt{HumanEval-NFR/51}}\\
        \multicolumn{4}{l}{Description: \textit{remove\_vowels is a function that takes string and returns string without vowels.}}\\
        \hdashline
        \small{General} & \small{The function takes a string `text' as input.} & \small{Correct} & \small{The function takes string with or without vowels as input.} \\
        \small{General} & \small{It returns a string.} & \small{Correct} & \small{The function should return a string without vowels.} \\
        \small{General} & \small{The function should remove all vowels from the input `text' and return the modified string.} & \small{Correct} & \small{The function should return a string without vowels from the input string.} \\
        \small{Edge} & \small{Handle an empty `text' string, returning an empty string.} & \small{Correct} & \small{Input can be empty string. In this case, the function should return empty string.} \\
        \small{Edge} & \small{Handle the case where `text' contains only vowels, returning an empty string.} & \small{Correct} & \small{Input can contain only vowels. In this case, all of vowels should be removed and the function should return empty string.} \\
        \small{Edge} & \small{Handle the case where `text' contains no vowels, returning the original string.} & \small{Correct} & \small{Input can contain no vowels. In this case, there are nothing to be removed and the function should return original string.} \\
        \small{Edge} & \small{Handle both lowercase and uppercase vowels.} & \small{Correct} & \small{Input can be string in lowercase and uppercase. Both of them should be handled.} \\
        \small{Performance} & \small{Time complexity: O(n) where n is the length of the `text' string.} & \small{Correct} & \small{To traverse the string once and check whether each character is a vowel or not, and then add non-vowel characters to the result string, it requires 
        $O(n)$ time, where $n$ is the length of the string.} \\
        \small{Performance} & \small{Space complexity: $O(n)$ where n is the length of the `text' string.} & \small{Correct} & \small{Creating a new string to store the non-vowel characters requires additional space. If the length of the input string is n, the length of the resulting string can be at most n, leading to an $O(n)$ space complexity.} \\
        \small{Performance} & \small{Ensure efficiency even for extremely large inputs, providing results within 5 seconds.} & \small{Correct} & \small{The inputs should be handled in predetermined testing time.} \\
        \small{Robustness} & \small{If a non-string input is provided, print an error message to `stderr' and return None.} & \small{Correct} & \small{Invalid inputs should be handled and return None.} \\
        \small{Maintainability} & \small{Target Cyclomatic Complexity: ≤ 5.} & \small{Correct} & \small{5 indicates low risk, where the code mainly consists of simple blocks.} \\
        \thickhline
    \end{tabularx}
    \caption{{Detailed explanation for the human validation of generated requirements by \ours~(Index: 0 in Table~\ref{tab:gen_reqs_analysis}) on HumanEval-NFR/51. } }
    \label{tab:gen_req_judgement_0}
\end{table*}

\begin{table*}[h]
    
    \centering
    \begin{tabularx}{\textwidth}{lXcX}
        \thickhline
        \multirow{2}{*}{\small{Sub-requirement}} & \multirow{2}{*}{\small{Generated Requirement}} & \small{Human} & \multirow{2}{*}{\small{Explanation}}  \\
        & & \small{Validation} & \\\hline
        \multicolumn{4}{c}{\texttt{HumanEval-NFR/11}}\\
        \multicolumn{4}{l}{{Description: \textit{Input are two strings a and b consisting only of 1s and 0s. }}}\\ 
        \multicolumn{4}{l}{{\phantom{Description: }\textit{Perform binary XOR on these inputs and return result also as a string.}}}\\
        \hdashline
        \small{General} & \small{The function takes two strings `a' and `b' as input.} & \small{Correct} & \small{The function should take two strings consisting only of 1s and 0s.} \\
        \small{General} & \small{It returns a string.} & \small{Correct} & \small{The function should return string having result of a binary XOR operation on these inputs.} \\
        \small{General} & \small{The function should perform a binary XOR operation on the input strings `a' and `b' and return the result as a string.} & \small{Correct} & \small{The function takes two binary strings and return a result of a binary XOR operation.} \\
        \small{Edge} & \small{Handle empty strings as input, returning an empty string.} & \small{Correct} & \small{The function should handle empty strings and return empty string.} \\
        \small{Edge} & \small{Handle strings with different lengths, returning None.} & \small{Correct} & \small{When the two inputs have different length, a binary XOR operation is impossible.} \\
        \small{Edge} & \small{Handle strings with characters other than `0' and `1', returning None.} & \small{Correct} & \small{When the two inputs are not valid binary string, a binary XOR operation is impossible.} \\
        \small{Performance} & \small{Time complexity: $O(n)$ where $n$ is the length of the longer input string.} & \small{Correct} & \small{The time complexity of performing binary XOR on two strings is proportional to the length of the strings.} \\
        \small{Performance} & \small{Space complexity: $O(n)$ where $n$ is the length of the longer input string.} & \small{Correct} & \small{If a new string is created to store the result, its length would be equal to the length of the input strings. } \\
        \small{Performance} & \small{Ensure efficiency even for extremely large inputs, providing results within 5 seconds.} & \small{Correct} & \small{The inputs should be handled in predetermined testing time.} \\
        \small{Robustness} & \small{If non-string inputs are provided, print an error message to `stderr' and return None.} & \small{Correct} & \small{Invalid inputs should be handled and return None.} \\
        \small{Robustness} & \small{If the input strings contain characters other than `0' and `1', print an error message to `stderr' and return None.} & \small{Correct} & \small{Invalid inputs should be handled and return None.} \\
        \small{Maintainability} & \small{Target Cyclomatic Complexity: ≤ 10.} & \small{Correct} & \small{10 indicates low risk, where the code consists of well-structured and stable blocks.} \\
        \thickhline
    \end{tabularx}
    \caption{{Detailed explanation for the human validation of generated requirements by \ours~(Index: 0 in Table~\ref{tab:gen_reqs_analysis}) on HumanEval-NFR/11. } }
    \label{tab:gen_req_judgement_1}
\end{table*}

\begin{table*}[h]
    
    \centering
    \begin{tabularx}{\textwidth}{lXcX}
        \thickhline
        \multirow{2}{*}{\small{Sub-requirement}} & \multirow{2}{*}{\small{Generated Requirement}} & \small{Human} & \multirow{2}{*}{\small{Explanation}}  \\
        & & \small{Validation} & \\\hline
        \multicolumn{4}{c}{\texttt{HumanEval-NFR/0}}\\
        \multicolumn{4}{l}{Description: \textit{Check if in given list of numbers, are any two numbers closer to each other than given}}\\
        \multicolumn{4}{l}{\phantom{Description: }\textit{threshold.}}\\
        \hdashline
        \small{General} & \small{The function takes a list of floating-point numbers `numbers' and a threshold value `threshold' as input.} & \small{Correct} & \small{The function takes two inputs. The first is a list of floating-point numbers and the second is a threshold value.} \\
        \small{General} & \small{It returns a boolean value.} & \small{Correct} & \small{The function should return True if there are any two numbers in the `numbers' list that are closer to each other than the given `threshold'. Otherwise, it should return False.} \\
        \small{General} & \small{The function should return `True' if there are any two numbers in the `numbers' list that are closer to each other than the given `threshold'. Otherwise, it should return `False'.} & \small{Correct} & \small{The function checks if there are any two numbers closer to each other than given threshold in given list of numbers.} \\
        \small{Edge} & \small{Handle an empty `numbers' list, returning `False'.} & \small{Correct} & \small{The function should return False if there are not any numbers in the `numbers' list.} \\
        \small{Edge} & \small{Handle the case where no two numbers in `numbers' are closer to each other than the `threshold', returning `False'.} & \small{Correct} & \small{The function should return True if there are any two numbers in the `numbers' list that are closer to each other than the given `threshold'. Otherwise, it should return False.} \\
        \small{Edge} & \small{Handle the case where all numbers in `numbers' are closer to each other than the `threshold', returning `True'.} & \small{Correct} & \small{The function should return True if there are any two numbers in the `numbers' list that are closer to each other than the given `threshold'. Otherwise, it should return False.} \\
        \small{Performance} & \small{Time complexity: $O(n^2)$ where $n$ is the length of the `numbers' list.} & \small{Correct} & \small{All elements of the given list need to be iterated through, and the difference between each pair of numbers needs to be calculated. This process is proportional to the size of the input list. Therefore, if we denote the length of the input list as $n$, the time complexity is $O(n^2)$.} \\
        \small{Performance} & \small{Space complexity: $O(1)$.} & \small{Correct} & \small{If the problem is solved without using additional memory, the space complexity is $O(1)$.} \\
        \small{Performance} & \small{Ensure efficiency even for extremely large inputs, providing results within 5 seconds.} & \small{Correct} & \small{The inputs should be handled in predetermined testing time.} \\
        \small{Robustness} & \small{If non-list `numbers' input or a non-float `threshold' input is provided, print an error message to `stderr' and return None.} & \small{Correct} & \small{Invalid inputs should be handled and return None.} \\
        \small{Robustness} & \small{If non-float elements in the `numbers' list or a negative `threshold' is provided, print an error message to `stderr' and return None.} & \small{Correct} & \small{Invalid inputs should be handled and return None.} \\
        \small{Maintainability} & \small{Target Cyclomatic Complexity: ≤ 10.} & \small{Correct} & \small{10 indicates low risk, where the code consists of well-structured and stable blocks.} \\
        \thickhline
    \end{tabularx}
    \caption{{Detailed explanation for the human validation of generated requirements by \ours~(Index: 0 in Table~\ref{tab:gen_reqs_analysis}) on HumanEval-NFR/0. } }
    \label{tab:gen_req_judgement_2}
\end{table*}

\begin{table*}[h]
    
    \centering
    \begin{tabularx}{\textwidth}{lXccc}
        \thickhline
        \multirow{2}{*}{\small{Index}} & \multirow{2}{*}{\small{Generated Test Cases}}  & \small{Sub-requirement} & \small{\# Passed} & \small{Human}  \\
        & & \small{Type} & \small{Cands.} & \small{Validation} \\\hline
        \multicolumn{4}{c}{\texttt{HumanEval-NFR/51}}\\\hdashline
        \small{0} & \footnotesize{assert remove\_vowels(`') == `', `Failed to handle an empty input string.'} & \small{General} & \small{10/10} & \small{Correct}\\
        \small{1} & \footnotesize{assert remove\_vowels("abcdef\textbackslash nghijklm") == `bcdf\textbackslash nghjklm', `Failed to remove all vowels from the input string.'} & \small{General} & \small{10/10} & \small{Correct}\\
        \small{2} & \footnotesize{assert remove\_vowels(`abcdef') == `bcdf', `Failed to remove vowels from the input string.'} & \small{General} & \small{10/10} & \small{Correct}\\
        \small{3} & \footnotesize{assert remove\_vowels(`aaaaa') == `', `Failed to handle case where the input string contains only vowels.'} & \small{Edge} & \small{10/10} & \small{Correct}\\
        \small{4} & \footnotesize{assert remove\_vowels(`aaBAA') == `B', `Failed to remove only lowercase vowels from the input string.'} & \small{Edge} & \small{10/10} & \small{Correct}\\
        \small{5} & \footnotesize{assert remove\_vowels(`zbcd') == `zbcd', `Failed to handle case where the input string contains no vowels.'} & \small{Edge} & \small{10/10} & \small{Correct}\\
        \small{6} & \footnotesize{assert remove\_vowels(`a' * 10**6) == `', `Failed to handle large input size.'} & \small{Performance} & \small{10/10} & \small{Correct}\\
        \small{7} & \footnotesize{assert remove\_vowels(`z' * 10**6) == `z' * 10**6, `Failed to handle large input size.'} & \small{Performance} & \small{10/10} & \small{Correct}\\
        \small{8} & \footnotesize{assert remove\_vowels(123) == None, `Failed to handle case where the input is not a string.'} & \small{Robustness} & \small{10/10} & \small{Correct}\\
        \small{9} & \footnotesize{assert result.total\_complexity <= 5, `Failed to have a Cyclomatic Complexity less than or equal to 5 by Radon.'} & \small{Maintainability} & \small{10/10} & \small{Correct}\\
        \hline
        \multicolumn{4}{c}{\texttt{HumanEval-NFR/83}}\\\hdashline 
        \small{0} & \footnotesize{assert starts\_one\_ends(1) == 9, `Failed to count the numbers of 1-digit positive integers that start or end with 1.'} & \small{General} & \small{0/10} & \small{Incorrect}\\
        \small{1} & \footnotesize{assert starts\_one\_ends(2) == 19, `Failed to count the numbers of 2-digit positive integers that start or end with 1.'} & \small{General} & \small{0/10} & \small{Correct}\\
        \small{2} & \footnotesize{assert starts\_one\_ends(3) == 271, `Failed to count the numbers of 3-digit positive integers that start or end with 1.'} & \small{General} & \small{0/10} & \small{Incorrect}\\
        \small{3} & \footnotesize{assert starts\_one\_ends(0) == 0, `Failed to handle the case where n is 0.'} & \small{Edge} & \small{0/10} & \small{Incorrect}\\
        \small{4} & \footnotesize{assert starts\_one\_ends(-5) == 0, `Failed to handle the case where n is negative.'} & \small{Edge} & \small{0/10} & \small{Incorrect}\\
        \small{5} & \footnotesize{assert starts\_one\_ends(10**6) == 900000, `Failed to handle large input size.'} & \small{Performance} & \small{0/10} & \small{Incorrect}\\
        \small{6} & \footnotesize{assert starts\_one\_ends(`invalid') == None, `Failed to handle case where the input n is not an integer.'} & \small{Robustness} & \small{10/10} & \small{Correct}\\
        \small{7} & \footnotesize{assert result.total\_complexity <= 5, `Failed to have a Cyclomatic Complexity less than or equal to 5 by Radon.'} & \small{Maintainability} & \small{2/10} & \small{Correct}\\
        \hline
        \multicolumn{4}{c}{\texttt{HumanEval-NFR/100}}\\\hdashline 
        \small{0} & \footnotesize{assert make\_a\_pile(3) == [3, 5, 7], `Failed to create the pile correctly.'} & \small{General} & \small{5/10} & \small{Correct}\\
        \small{1} & \footnotesize{assert make\_a\_pile(5) == [5, 7, 9, 11, 13], `Failed to create the pile correctly.'} & \small{General} & \small{5/10} & \small{Correct}\\
        \small{2} & \footnotesize{assert make\_a\_pile(0) == [], `Failed to handle the case where n is 0.'} & \small{Edge} & \small{5/10} & \small{Correct}\\
        \small{3} & \footnotesize{assert make\_a\_pile(-5) == None, `Failed to handle the case where n is negative.'} & \small{Edge} & \small{5/10} & \small{Correct}\\
        \small{4} & \footnotesize{assert make\_a\_pile(10**6) == list(range(10**6, 10**6 + 2 * 10**6, 2)), `Failed to handle large input size.'} & \small{Performance} & \small{0/10} & \small{Correct}\\
        \small{5} & \footnotesize{assert make\_a\_pile(`invalid') == None, `Failed to handle the case where the input n is not an integer.'} & \small{Robustness} & \small{5/10} & \small{Correct}\\
        \small{6} & \footnotesize{assert make\_a\_pile(0.5) == None, `Failed to handle the case where the input n is not a positive integer.'} & \small{Robustness} & \small{5/10} & \small{Correct}\\
        \small{7} & \footnotesize{assert result.total\_complexity <= 10, `Failed to have a Cyclomatic Complexity less than or equal to 10 by Radon.'} & \small{Maintainability} & \small{10/10} & \small{Correct}\\
        \thickhline
    \end{tabularx}
    \caption{{Validation of generated test cases by \ours~on HumanEval-NFR.} }
    \label{tab:gen_tcs_analysis}
\end{table*}

\begin{table*}[h]
    \centering

    \begin{tabularx}{\textwidth}{X|X}
        \thickhline
        \multicolumn{2}{c}{Preferred NFR(s)} \\\hline
        \textit{Time Perf.}  & \textit{All - Time Perf.}  \\\hdashline
    \end{tabularx}
    \renewcommand{\arraystretch}{0.8}
    \begin{tabularx}{\textwidth}{X|X}
{\small} & {\small \# Problem Agnostic Requirements} \\
{\small} &{\small - Ensure error-free execution, avoiding type errors, arithmetic operation errors (e.g., division by zero, overflow), and so on. } \\
{\small} &{\small - Demonstrate efficiency in the algorithmic approach to problem-solving. } \\
{\small} &{\small - Code should be clear, well-commented, and easy to understand and maintain. } \\
{\small} &{\small} \\
{\small \# Functional Requirements } & {\small \# Functional Requirements } \\

{\small \#\# Input-output Conditions } & {\small \#\# Input-output Conditions } \\

{\small - The function takes an integer `decimal` as input. } & {\small - The function takes an integer `decimal` as input. } \\

{\small - It returns a string. } &  {\small - It returns a string. } \\

{\small } & {\small} \\

{\small \#\# Expected Behavior } & {\small \#\# Expected Behavior } \\

{\small - The function should convert the given decimal number to binary and return a string representation of the binary number. } & {\small - The function should convert the decimal number to binary and return a string representation of the binary number. } \\

{\small - The resulting string should have the characters `db' at the beginning and end. } & {\small - The string should have the format "db" + binary representation + "db". } \\
{\small } &  {\small - Each character in the binary representation should be `0' or `1’. } \\

{\small } & {\small} \\

{\small \#\# Edge Cases } & {\small \#\# Edge Cases } \\

{\small - Handle negative decimal numbers, returning the binary representation with a `-' sign. } & {\small - Handle the case where `decimal` is 0, returning "db0db". } \\

{\small - Handle decimal numbers equal to 0, returning the string "db0db". } & {\small - Handle the case where `decimal` is negative, returning None. } \\
{\small } & {\small - Handle the case where `decimal` is not an integer, returning None. } \\

{\small } & {\small} \\
{\small \# Non-functional Requirements } & {\small \# Non-functional Requirements } \\
{\small} & {\small \#\# Specific Quality Requirements } \\
{\small \#\# Performance } & {\small} \\

{\small - Time complexity: O(log n) where n is the value of the `decimal` input. } & {\small} \\
{\small - Space complexity: O(log n) where n is the value of the `decimal` input. } & {\small} \\
{\small - Ensure efficiency even for extremely large inputs, providing results within 5 seconds. } & {\small} \\

{\small} & {\small \#\#\# Robustness } \\
{\small} &{\small - If a non-integer `decimal` input is provided, print an error message to `stderr` and return None. } \\
{\small} &{\small - If a negative `decimal` input is provided, print an error message to `stderr` and return None. } \\
{\small} &{\small} \\
{\small} &{\small \#\#\# Maintainability } \\
{\small} &{\small - Target Cyclomatic Complexity ≤ 5. } \\
        \thickhline
    \end{tabularx}
    \caption{Real life examples of generated requirements for \texttt{HumanEval-NFR/79} by \ours~with Plug-and-Play NFR Preference Control. } 
    \label{tab:case_study_nfr_tradeoff}
\end{table*}

\begin{table*}[h]
    \centering

    \begin{tabularx}{\textwidth}{X|X}
        \thickhline
        \multicolumn{2}{c}{Preferred NFR(s)} \\\hline
        \textit{Time Perf.}  & \textit{All - Time Perf.}  \\\hdashline
    \end{tabularx}
    \renewcommand{\arraystretch}{0.8}
    \begin{tabularx}{\textwidth}{X|X}
{\small import sys} & {\small import sys} \\

{\small import math } & {\small import math } \\

{\small import re } &{\small import re } \\

{\small import numpy } &{\small import numpy } \\

{\small import numpy as np } &{\small import numpy as np } \\

{\small from typing import * } &{\small from typing import * } \\

{\small } &{\small } \\

{\small def decimal\_to\_binary(decimal): } &{\small def decimal\_to\_binary(decimal): } \\

{\small$\quad$binary = bin(decimal)[2:] \# Convert decimal to binary and remove `0b' prefix } &{\small $\quad$\# Input Validation } \\

{\small $\quad$return `db' + binary + `db' \# Add `db' prefix and suffix to the binary representation } &{\small $\quad$if not isinstance(decimal, int): } \\
& {\small $\quad$$\quad$sys.stderr.write('Invalid input: decimal must be an integer.’) } \\
&{\small $\quad$return None } \\
&{\small } \\
&{\small $\quad$\# Initialize variables } \\
&{\small $\quad$binary = "" } \\
&{\small } \\
&{\small $\quad$\# Handle special case for 0 } \\
&{\small $\quad$if decimal == 0: } \\
&{\small $\quad$$\quad$return "db0db" } \\
&{\small } \\
&{\small $\quad$\# Convert decimal to binary } \\
&{\small $\quad$while decimal > 0: } \\
&{\small $\quad$$\quad$binary += str(decimal \% 2) } \\
&{\small $\quad$$\quad$decimal //= 2 } \\
&{\small } \\
&{\small $\quad$\# Reverse the binary string } \\
&{\small $\quad$binary = binary[::-1] } \\
&{\small } \\
&{\small $\quad$\# Add prefix and suffix } \\
&{\small $\quad$binary = "db" + binary + "db" } \\
&{\small } \\
&{\small $\quad$return binary } \\
        \thickhline
    \end{tabularx}
    \caption{Real life examples of generated code snippets for \texttt{HumanEval-NFR/79} by \ours~with Plug-and-Play NFR Preference Control. The conditioned requirements are shown in Table~\ref{tab:case_study_nfr_tradeoff}. } 
    \label{tab:case_study_nfr_tradeoff_code}
\end{table*}

\newpage
\begin{table*}[t]
\section{In-Context Learning Examples}
\begin{tabularx}{\textwidth}{X}
\thickhline
\textbf{Problem Description Example}\\
\hdashline
from typing import List\\
\\
def longest\_subarray\_with\_sum\_limit(nums: List[int], target: int) -> List[int]:\\
$\quad$"""Find the longest subarray of the given array `nums' such that the sum of its elements is less than or equal to the given `target'.\\
$\quad$$\quad$  \text{>}\text{>}\text{>} longest\_subarray\_with\_sum\_limit([3, 1, 5, 2, 1, 9, 7, 4], 10)\\
$\quad$$\quad$  [3, 1, 5]\\
$\quad$$\quad$  \text{>}\text{>}\text{>} longest\_subarray\_with\_sum\_limit([1, 2, 3, 4, 5], 3)\\
$\quad$$\quad$  [1, 2]\\
$\quad$"""\\
\hline
\textbf{Requirement Generation Prompt}\\
\hdashline
Write requirements for the problem.\\
\hline
\textbf{Requirement Generation Example}\\
\hdashline
\end{tabularx}
\renewcommand{\arraystretch}{0.7}
\begin{tabularx}{\textwidth}{X}
{\small
\# Problem Agnostic Requirements}\\{\small
- Ensure error-free execution, avoiding type errors, index errors, arithmetic operation errors (e.g., division by zero, overflow), and so on.}\\{\small
- Demonstrate efficiency in the algorithmic approach to problem-solving.}\\{\small
- Code should be clear, well-commented, and easy to understand and maintain.}\\{\small
}\\{\small
\# Functional Requirements}\\{\small
\#\# Input-output Conditions}\\{\small
- The function takes a list of integers `nums' and a target integer `target' as input.}\\{\small
- It returns a list of integers.}\\{\small
}\\{\small
\#\# Expected Behavior}\\{\small
- The function should return the longest subarray of `nums' whose sum is less than or equal to `target'.}\\{\small
}\\{\small
\#\# Edge Cases}\\{\small
- Handle an empty `nums' list, returning an empty list.}\\{\small
- Handle the case where no subarray in `nums' satisfies the condition, returning an empty list.}\\{\small
- Handle multiple subarrays with the same maximum length, returning any of them.}\\{\small
}\\{\small
\# Non-functional Requirements}\\{\small
\#\# Performance}\\{\small
- Time complexity: O(n) where n is the length of the `nums' list.}\\{\small
- Space complexity: O(1).}\\{\small
- Ensure efficiency even for extremely large inputs, providing results within 5 seconds.}\\{\small
}\\{\small
\#\# Specific Quality Requirements}\\{\small
\#\#\# Robustness}\\{\small
- If non-list `nums' input or a non-integer `target' input is provided, print an error message to `stderr' and return None.}\\{\small
- If non-integer elements in the `nums' list or a negative `target' is provided, print an error message to `stderr' and return None.}\\{\small
}\\{\small
\#\#\# Reliability}\\{\small
- Avoid index errors while accessing `nums' elements.}\\{\small
}\\{\small
\#\#\# Maintainability}\\{\small
- Target Cyclomatic Complexity ≤ 10.}\\
\thickhline
\end{tabularx}
\caption{The in-context example for requirement generation by \ours. The problem description example is from~\citet{shinn2023reflexion}.}
\label{tab:ice_rq_gen}
\end{table*}

\begin{table*}[h]
\begin{tabularx}{\textwidth}{X}
\thickhline
\textbf{Problem Description Example}\\
\hdashline
from typing import List ...\\
\hline
\textbf{Requirement Generation Prompt}\\
\hdashline
Write requirements for the problem.\\
\hline
\textbf{Requirement Generation Example}\\
\hdashline
\# Problem Agnostic Requirements ...\\
\hline
\textbf{Code Generation Prompt}\\
\hdashline
Write the code for the problem.\\
\hline
\textbf{Code Generation Example}\\
\hdashline
\end{tabularx}
\renewcommand{\arraystretch}{0.8}
\begin{tabularx}{\textwidth}{X}
from typing import List, Optional\\import sys\\\\def longest\_subarray\_with\_sum\_limit(nums: Optional[List[int]], target: Optional[int]) -> List[int]:\\$\quad$\# Validate inputs\\$\quad$if not isinstance(nums, list) or not all(isinstance(num, int) for num in nums):\\$\quad$$\quad$sys.stderr.write('Invalid input: nums must be a list of integers.')\\$\quad$$\quad$return\\$\quad$if not isinstance(target, int):\\$\quad$$\quad$sys.stderr.write('Invalid input: target must be an integer.')\\$\quad$$\quad$return\\\\$\quad$n = len(nums)\\$\quad$left, right = 0, 0\\$\quad$max\_length = 0\\$\quad$current\_sum = 0\\$\quad$result = []\\$\quad$\\$\quad$\# Handle empty list\\$\quad$if n == 0:\\$\quad$$\quad$return []\\\\$\quad$\# Initial window expansion\\$\quad$while right < n and current\_sum + nums[right] <= target:\\$\quad$$\quad$current\_sum += nums[right]\\$\quad$$\quad$right += 1\\$\quad$\\$\quad$\# Update result if a valid window was found during initial expansion\\$\quad$if right > 0:\\$\quad$$\quad$result = nums[left:right]\\$\quad$$\quad$max\_length = right\\$\quad$\\$\quad$\# Main loop: Slide the window through the array\\$\quad$while right < n:\\$\quad$$\quad$current\_sum += nums[right]\\$\quad$$\quad$\\$\quad$$\quad$\# Shrink window from the left if current\_sum exceeds target\\$\quad$$\quad$while current\_sum > target:\\$\quad$$\quad$$\quad$current\_sum -= nums[left]\\$\quad$$\quad$$\quad$left += 1\\$\quad$$\quad$\\$\quad$$\quad$\# Update result if a longer valid window is found\\$\quad$$\quad$if right - left + 1 > max\_length:\\$\quad$$\quad$$\quad$max\_length = right - left + 1\\$\quad$$\quad$$\quad$result = nums[left:right+1]\\$\quad$$\quad$\\$\quad$$\quad$right += 1\\$\quad$\\$\quad$return result\\
\thickhline
\end{tabularx}
\caption{The in-context example for code generation by \ours. The skipped contents---Problem Description Example and Requirement Generation Example---are available in Table~\ref{tab:ice_rq_gen}.}
\label{tab:ice_code_gen}
\end{table*}

\begin{table*}[h]
\begin{tabularx}{\textwidth}{X}
\thickhline
\textbf{Problem Description Example}\\
\hdashline
from typing import List ...\\
\hline
\textbf{Requirement Generation Prompt}\\
\hdashline
Write requirements for the problem.\\
\hline
\textbf{Requirement Generation Example}\\
\hdashline
\# Problem Agnostic Requirements ...\\
\hline
\textbf{Test Case Generation Prompt}\\
\hdashline
Write test cases for the problem.\\
\hline
\textbf{Test Case Generation Example}\\
\hdashline
\end{tabularx}
\renewcommand{\arraystretch}{0.56}
\begin{tabularx}{\textwidth}{X}
{\tiny
\# Test Cases Regarding Functional Requirements}\\{\tiny
\#\# General Cases}\\{\tiny
\#\#\# The longest subarray with sum less than or equal to 10 is [1, 5, 2, 1]}\\{\tiny
assert longest\_subarray\_with\_sum\_limit([3, 1, 5, 2, 1, 9, 7, 4], 10) == [1, 5, 2, 1], `Failed to find the longest subarray.'}\\{\tiny
}\\{\tiny
\#\#\# None of the subarrays have a sum less than or equal to 3}\\{\tiny
\#\#\# The function should return an empty list}\\{\tiny
assert longest\_subarray\_with\_sum\_limit([2, 3, 4, 5, 6], 1) == [], `Failed to handle case where no subarray satisfies the condition.'}\\{\tiny
}\\{\tiny
\#\# Edge Cases}\\{\tiny
\#\#\# The input list is empty, so the function should return an empty list}\\{\tiny
assert longest\_subarray\_with\_sum\_limit([], 5) == [], `Failed to handle an empty input list.'}\\{\tiny
}\\{\tiny
\#\#\# The longest subarray with sum less than or equal to 10 is [5, 5]}\\{\tiny
assert longest\_subarray\_with\_sum\_limit([5, 5, 5, 5], 10) == [5, 5], `Failed to find the longest subarray when all elements have the same value.'}\\{\tiny
}\\{\tiny
\#\#\# The entire nums array is a valid subarray with sum less than or equal to 15}\\{\tiny
assert longest\_subarray\_with\_sum\_limit([1, 2, 3, 4, 5], 15) == [1, 2, 3, 4, 5], `Failed to handle case where the entire array is a valid subarray.'}\\{\tiny
}\\{\tiny
\# Test Cases Regarding Non-functional Requirements}\\{\tiny
\#\# Performance Requirements}\\{\tiny
\#\#\# The nums list contains $10^6$ elements with increasing values from 1 to $10^6$}\\{\tiny
\#\#\# The longest subarray with sum less than or equal to $10^5$ is [1, 2, 3, ..., $10^5$]}\\{\tiny
assert longest\_subarray\_with\_sum\_limit(list(range(1, 10**6 + 1)), 10**5) == list(range(1, 446 + 1)), `Failed to handle large input size.'}\\{\tiny
}\\{\tiny
\#\#\# The nums list contains $10^6$ elements with increasing values from 1 to $10^6$}\\{\tiny
\#\#\# The longest subarray with sum less than or equal to 2 is [1]}\\{\tiny
assert longest\_subarray\_with\_sum\_limit(list(range(1, 10**6 + 1)), 2) == [1], `Failed to handle case where the subarray length is 1.'}\\{\tiny
}\\{\tiny
\#\#\# The nums list contains $10^6$ elements with increasing values from 1 to $10^6$}\\{\tiny
\#\#\# The longest subarray with sum less than or equal to 5 is [1, 2]}\\{\tiny
assert longest\_subarray\_with\_sum\_limit(list(range(1, 10**6 + 1)), 5) == [1, 2], `Failed to handle case where the subarray length is 2.'}\\{\tiny
}\\{\tiny
\#\#\# The nums list contains $10^6$ elements, all of which are $10^6$}\\{\tiny
\#\#\# The sum of all the elements in nums is $10^12$, which is larger than the target}\\{\tiny
\#\#\# Therefore, there is no subarray that satisfies the condition and the function should return an empty list}\\{\tiny
assert longest\_subarray\_with\_sum\_limit([10**8] * 10**6, 10**7) == [], `Failed to handle case where no subarray satisfies the condition.'}\\{\tiny
}\\{\tiny
\#\#\# The nums list contains $10^6$ elements with increasing values from 1 to $10^6$}\\{\tiny
\#\#\# The target is the sum of all the elements in nums, so the entire array is a valid subarray}\\{\tiny
assert longest\_subarray\_with\_sum\_limit(list(range(1, 10**5 + 1)), 10**10) == list(range(1, 10**5 + 1)), `Failed to handle case where the entire array is a valid subarray.'}\\{\tiny
}\\{\tiny
\#\# Specific Quality Requirements}\\{\tiny
\#\#\# Robustness}\\{\tiny
\#\#\#\# The sum of the elements in nums is larger than the target,}\\{\tiny
\#\#\#\# so the function should return None}\\{\tiny
assert longest\_subarray\_with\_sum\_limit([10**20, 10**20], 10**19) == None, `Failed to handle case where the sum of the elements in nums is larger than the target.'}\\{\tiny
}\\{\tiny
\#\#\#\# The nums input is not a list of integers, so the function should return None}\\{\tiny
assert longest\_subarray\_with\_sum\_limit('invalid', 10) == None, `Failed to handle case where the input nums is not a list.'}\\{\tiny
}\\{\tiny
\#\#\#\# The target input is not an integer, so the function should return None}\\{\tiny
assert longest\_subarray\_with\_sum\_limit([1, 2, 3], `invalid') == None, `Failed to handle case where the input target is not an integer.'}\\{\tiny
}\\{\tiny
\#\#\#\# The nums list contains elements that are not integers, so the function should return None}\\{\tiny
assert longest\_subarray\_with\_sum\_limit([1, 2, `invalid', 4], 5) == None, `Failed to handle case where the input nums contains non-integer elements.'}\\{\tiny
}\\{\tiny
\#\#\#\# The target is a negative number, so the function should return None}\\{\tiny
assert longest\_subarray\_with\_sum\_limit([1, 2, 3, 4, 5], -10) == None, `Failed to handle case where the input target is negative.'}\\{\tiny
}\\{\tiny
\#\#\# Reliability}\\{\tiny
\#\#\#\# Satisfied if no errors occur across all test cases}\\{\tiny
}\\{\tiny
\#\#\# Maintainability}\\{\tiny
\#\#\#\# Calculate Cyclomatic Complexity using Radon}\\{\tiny
\#\#\#\# Check if the Cyclomatic Complexity is less than or equal to 10}\\{\tiny
from radon.visitors import ComplexityVisitor}\\{\tiny
result = ComplexityVisitor.from\_code("""\textbf{\$\{Generated Code\}}""")}\\{\tiny
assert result.total\_complexity <= 10, `Failed to have a Cyclomatic Complexity less than or equal to 10 by Radon.'}\\
\thickhline
\end{tabularx}
\caption{The in-context example for test case generation by \ours. The skipped contents---Problem Description Example and Requirement Generation Example---are available in Table~\ref{tab:ice_rq_gen}. \textbf{\$\{Generated Code\}} denotes the string text of the code that is to be checked. }
\label{tab:ice_tc_gen}
\end{table*}

\begin{table}[t]
\section{In-Context Learning Prompt Templates}
\begin{tabularx}{0.48\textwidth}{X}
\thickhline
\textbf{\{Problem Description Example\}}\\
Write requirements for the problem.\\
\textbf{\{Requirement Generation Example\}}\\
\\
\textbf{\{Test Problem Description\}}\\
Write requirements for the problem.\\
\thickhline
\end{tabularx}
\caption{The in-context learning prompt template for requirement generation by \ours.}
\label{tab:icl_rq_gen_prompt}
\end{table}

\begin{table}[th]
\begin{tabularx}{0.48\textwidth}{X}
\thickhline
\textbf{\{Problem Description Example\}}\\
Write requirements for the problem.\\
\textbf{\{Requirement Generation Example\}}\\
Write the code for the problem.\\
\textbf{\{Code Generation Example\}}\\
\\
\textbf{\{Test Problem Description\}}\\
Write requirements for the problem.\\
\textbf{\{Generated Requirements by \textsc{ArchCode}\}}\\
Write the code for the problem.\\
\thickhline
\end{tabularx}
\caption{The in-context learning prompt template for code generation by \ours.}
\label{tab:icl_code_gen_prompt}
\end{table}

\begin{table}[t]
\begin{tabularx}{0.48\textwidth}{X}
\thickhline
\textbf{\{Problem Description Example\}}\\
Write requirements for the problem.\\
\textbf{\{Requirement Generation Example\}}\\
Write test cases for the problem.\\
\textbf{\{Test Case Generation Example\}}\\
\\
\textbf{\{Test Problem Description\}}\\
Write requirements for the problem.\\
\textbf{\{Generated Requirements by \textsc{ArchCode}\}}\\
Write test cases for the problem.\\
\thickhline
\end{tabularx}
\caption{The in-context learning prompt template for test case generation by \ours.}
\label{tab:icl_tc_gen_prompt}
\end{table}

\begin{table}[th]
\begin{tabularx}{0.48\textwidth}{X}
\thickhline
\textbf{\{Problem Description Example\}}\\
Write the code for the problem.\\
\textbf{\{Code Generation Example\}}\\
\\
\textbf{\{Test Problem Description\}}\\
Write the code for the problem.\\
\thickhline
\end{tabularx}
\caption{The in-context learning prompt template for code generation by GPT-3.5-Turbo baseline.}
\label{tab:icl_code_gen_prompt_baseline}
\end{table}

\begin{table}[t]
\begin{tabularx}{0.48\textwidth}{X}
\thickhline
\textbf{\{Problem Description Example\}}\\
{Write the plan for the problem.}\\
{\textbf{\{Chain-of-Thought Example\}}}\\
Write the code for the problem.\\
\textbf{\{Code Generation Example\}}\\
\\
\textbf{\{Test Problem Description\}}\\
Write the {plan} for the problem.\\
\thickhline
\end{tabularx}
\caption{The in-context learning prompt template for code generation by GPT-3.5-Turbo + CoT baseline.}
\label{tab:icl_code_gen_prompt_cot_baseline}
\end{table}

\begin{table*}[t]
\section{Case Study}
\label{appendix:case_study}
\begin{tabularx}{\textwidth}{X}
\thickhline
\textbf{Generated Requirements}\\
\hdashline
\end{tabularx}
\renewcommand{\arraystretch}{0.7}
\begin{tabularx}{\textwidth}{X}
{\small
\# Problem Agnostic Requirements}\\{\small
- Ensure error-free execution, avoiding type errors, index errors, arithmetic operation errors (e.g., division by zero, overflow), and so on.}\\{\small
- Demonstrate efficiency in the algorithmic approach to problem-solving.}\\{\small
- Code should be clear, well-commented, and easy to understand and maintain.}\\{\small
}\\{\small
\# Functional Requirements}\\{\small
\#\# Input-output Conditions}\\{\small
- The function takes two integers `x` and `base` as input.}\\{\small
- It returns a string.}\\{\small
}\\{\small
\#\# Expected Behavior}\\{\small
- The function should convert the number `x` to the given `base` and return the string representation of the converted number.}\\{\small
}\\{\small
\#\# Edge Cases}\\{\small
- Handle the case where `x` is 0, returning '0'.}\\{\small
- Handle the case where `base` is 1, returning a string of '1' repeated `x` times.}\\{\small
- Handle the case where `base` is less than 2 or greater than 9, returning None.}\\{\small
- Handle the case where `x` is negative, returning None.}\\{\small
}\\{\small
\# Non-functional Requirements}\\{\small
\#\# Performance}\\{\small
- Time complexity: O(log(x)) where x is the value of `x`.}\\{\small
- Space complexity: O(log(x)) to store the result string.}\\{\small
- Ensure efficiency even for extremely large inputs, providing results within 5 seconds.}\\{\small
}\\{\small
\#\# Specific Quality Requirements}\\{\small
\#\#\# Robustness}\\{\small
- If a non-integer `x` or `base` input is provided, print an error message to `stderr` and return None.}\\{\small
- If a negative `base` is provided, print an error message to `stderr` and return None.}\\{\small
}\\{\small
\#\#\# Maintainability}\\{\small
- Target Cyclomatic Complexity: ≤ 10.}\\
\thickhline
\end{tabularx}
\caption{A real-life example of generated requirements for \texttt{HumanEval-NFR/44} by \ours.}
\label{tab:gen_rq}
\end{table*}

\begin{table*}[t]
\begin{tabularx}{\textwidth}{X}
\thickhline
\textbf{Generated Code}\\
\hdashline
\end{tabularx}
\renewcommand{\arraystretch}{0.8}
\begin{tabularx}{\textwidth}{X}
def change\_base(x: int, base: int) -> str:\\
$\quad$\# Input validation\\
$\quad$if not isinstance(x, int) or not isinstance(base, int):\\
$\quad$$\quad$sys.stderr.write('Invalid input: x and base must be integers.')\\
$\quad$$\quad$return None\\
$\quad$if base < 2 or base > 9:\\
$\quad$$\quad$sys.stderr.write('Invalid input: base must be between 2 and 9 (inclusive).')\\
$\quad$$\quad$return None\\
$\quad$if x < 0:\\
$\quad$$\quad$sys.stderr.write('Invalid input: x must be a non-negative integer.')\\
$\quad$$\quad$return None\\
$\quad$\\
$\quad$\# Handle special case\\
$\quad$if x == 0:\\
$\quad$$\quad$return '0'\\
$\quad$\\
$\quad$result = ''\\
$\quad$while x > 0:\\
$\quad$$\quad$remainder = x \% base\\
$\quad$$\quad$result = str(remainder) + result\\
$\quad$$\quad$x = x // base\\
$\quad$\\
$\quad$return result\\
\thickhline
\end{tabularx}
\caption{A real-life example of generated code for \texttt{HumanEval-NFR/44} by \ours.}
\label{tab:gen_code}
\end{table*}

\begin{table*}[t]
\begin{tabularx}{\textwidth}{X}
\thickhline
\textbf{Generated Test Cases}\\
\hdashline
\end{tabularx}
\renewcommand{\arraystretch}{0.56}
\begin{tabularx}{\textwidth}{X}
{\small
\# Test Cases Regarding Functional Requirements}\\{\small
\#\# General Cases}\\{\small
\#\#\# The number 8 in base 3 is represented as '22'}\\{\small
assert change\_base(8, 3) == '22', 'Failed to convert number to the desired base.'}\\{\small
}\\{\small
\#\#\# The number 8 in base 2 is represented as '1000'}\\{\small
assert change\_base(8, 2) == '1000', 'Failed to convert number to the desired base.'}\\{\small
}\\{\small
\#\#\# The number 7 in base 2 is represented as '111'}\\{\small
assert change\_base(7, 2) == '111', 'Failed to convert number to the desired base.'}\\{\small
}\\{\small
\#\# Edge Cases}\\{\small
\#\#\# The input number is 0, so the function should return '0' in any base}\\{\small
assert change\_base(0, 3) == '0', 'Failed to handle case where the input number is 0.'}\\{\small
}\\{\small
\#\#\# The input base is 1, which is not a valid base, so the function should return None}\\{\small
assert change\_base(8, 1) is None, 'Failed to handle case where the base is not a valid base.'}\\{\small
}\\{\small
\#\#\# The input number is negative, so the function should return None}\\{\small
assert change\_base(-8, 2) is None, 'Failed to handle case where the input number is negative.'}\\{\small
}\\{\small
\#\#\# The input base is greater than 9, which is not a valid base, so the function should return None}\\{\small
assert change\_base(8, 10) is None, 'Failed to handle case where the base is not a valid base.'}\\{\small
}\\{\small
\# Test Cases Regarding Non-functional Requirements}\\{\small
\#\# Performance Requirements}\\{\small
\#\#\# The input number is $10^6$, and the base is 2}\\{\small
\#\#\# The function should return a string representation of the number in base 2 within a reasonable time frame}\\{\small
assert change\_base(10**6, 2) == '11110100001001000000', 'Failed to handle large input size.'}\\{\small
}\\{\small
\#\#\# The input number is $10^6$, and the base is 9}\\{\small
\#\#\# The function should return a string representation of the number in base 9 within a reasonable time frame}\\{\small
assert change\_base(10**6, 9) == '1783661', 'Failed to handle large input size.'}\\{\small
}\\{\small
\#\#\# The input number is $10^18$, and the base is 2}\\{\small
\#\#\# The function should return a string representation of the number in base 2 within a reasonable time frame}\\{\small
assert change\_base(10**18, 2) == '110111100000101101101011001110100111011001000000000000000000', 'Failed to handle large input size.'}\\{\small
}\\{\small
\#\# Specific Quality Requirements}\\{\small
\#\#\# Robustness}\\{\small
\#\#\#\# The input number is not an integer, so the function should return None}\\{\small
assert change\_base('invalid', 2) is None, 'Failed to handle case where the input number is not an integer.'}\\{\small
}\\{\small
}\\{\small
\#\#\#\# The input base is not an integer, so the function should return None}\\{\small
assert change\_base(8, 'invalid') is None, 'Failed to handle case where the input base is not an integer.'}\\{\small
}\\{\small
\#\#\# Reliability}\\{\small
\#\#\#\# Satisfied if no errors occur across all test cases}\\{\small
}\\{\small
\#\#\# Maintainability}\\{\small
\#\#\#\# Calculate Cyclomatic Complexity using Radon}\\{\small
\#\#\#\# Check if the Cyclomatic Complexity is less than or equal to 10}\\{\small
import inspect}\\{\small
from radon.visitors import ComplexityVisitor}\\{\small
result = ComplexityVisitor.from\_code("""\textbf{\$\{Generated Code\}}""")}\\{\small
assert result.total\_complexity <= 10, `Failed to have a Cyclomatic Complexity less than or equal to 10 by Radon.'}\\
\thickhline
\end{tabularx}
\caption{A real-life example of generated test cases for \texttt{HumanEval-NFR/44} by \ours. \textbf{\$\{Generated Code\}} denotes the string text of the code that is to be checked. }
\label{tab:gen_tcs}
\end{table*}

\begin{table*}[t]

\begin{tabularx}{\textwidth}{X}
\thickhline
\textbf{Generated Requirements}\\
\hdashline
\end{tabularx}
\renewcommand{\arraystretch}{0.8}
\begin{tabularx}{\textwidth}{X}
\small    \textbf{\# Problem Agnostic Requirements} \\
\small    - Ensure execution is error-free, mitigating type errors, index errors, and arithmetic operation errors (e.g., division by zero, overflow) among others. \\
\small    - Showcase efficiency in the algorithmic approach to problem-solving. \\
\small    - Ensure code is clear, well-commented, and both easy to understand and maintain. \\
\small    \textbf{\# Functional Requirements} \\
\small    \textbf{\#\# Input-output Conditions} \\
\small    \textbf{\#\#\# Inputs} \\
\small    - Initial input values: $n$ and $m$ \\
\small      - Must be positioned in the first line of input \\
\small      - Adhere to the format: "$\{n\} \{m\}$" \\
\small      - Integer $n$ range: ($0 \leq n < 2000$) \\
\small      - Integer $m$ range: ($0 \leq m < 2000$) \\
\small    - Subsequent input values: grid denoting the positions of telephone poles \\
\small      - Each input line format: "$\{a_{i,1}\}\{a_{i,2}\}...\{a_{i,m+1}\}$" \\
\small      - Integer $a_{i,j}$ range: ($0 \leq a_{i,j} \leq 1$) \\
\small      - Each line represents a row in the grid \\
\small      - There is at least one telephone pole in the given grid. \\
\small    \textbf{\#\#\# Outputs} \\
\small    - Output must be a single integer denoting the value of $\sum_{x=0}^{n} \{\sum_{y=0}^{m} \{S(x, y)\}\}$ \\
\small    \textbf{\#\# Expected Behavior} \\
\small    - The city is represented as a plane. \\
\small    - The plane is represented by a grid of size $(n + 1) * (m + 1)$. \\
\small    - Each point $(x, y)$ on the plane has a square of the Euclidean distance between the nearest pole and $(x, y)$ denoted as $S(x, y)$. \\
\small    - $S(x, y)$ is defined as $(x_2 - x_1)^2 + (y_2 - y_1)^2$, where $(x_1, y_1)$ is the nearest pole and $(x_2, y_2)$ is $(x, y)$. \\
\small    - The code must calculate the sum of all $S(x, y)$ for each $0 \leq x \leq n$ and $0 \leq y \leq m$. \\
\small    - The code must output the value of $\sum_{x=0}^{n} \{\sum_{y=0}^{m} \{S(x, y)\}\}$ as a single integer. \\
\small    \textbf{\#\# Edge Cases} \\
\small    Ensure the code functions accurately, even if: \\
\small    - Inputs reach the condition boundaries \\
\small    - $n$ or $m$ is 0 \\
\small    - $n$ or $m$ is the maximum value \\
\small    - There is only one telephone pole in the grid \\
\small    \textbf{\# Non-functional Requirements} \\
\small    \textbf{\#\# Performance} \\
\small    - Test time limit: 2 seconds \\
\small    - Test memory limit: 256 megabytes \\
\small    \textbf{\#\# Specific Quality Requirements} \\
\small    \textbf{\#\#\# Robustness} \\
\small    Ensure the code operates correctly even if: \\
\small    - Input contains redundant spaces or newlines \\
\small    Enforce the code to print an appropriate error message if: \\
\small    - Input format is incorrect \\
\small    - Input data type is invalid \\
\small    - Input is out of the specified range \\
\small    - The grid does not contain at least one telephone pole. \\
\small    \textbf{\#\#\# Reliability} \\
\small    - Mitigate type errors during input and output stream management. \\
\small    - Prevent index errors while managing grid values. \\
\small    \textbf{\#\#\# Maintainability} \\
\small    - Aim for a Cyclomatic Complexity of: $\leq 10$.\\
\thickhline
\end{tabularx}
\caption{
    {A real-life example of generated requirements for \texttt{CodeContests/2} \texttt{(1575\_M. Managing Telephone Poles)} by \ours.}
}   
\end{table*}

\begin{table*}[t]

\begin{tabularx}{\textwidth}{X}
\thickhline
\textbf{Generated Code}\\
\hdashline
\end{tabularx}
\renewcommand{\arraystretch}{0.8}
\begin{tabularx}{\textwidth}{X}
\small    import sys\\
\small    import math\\
\small    \\
\small    def read\_line():\\
\small    \ \ \ \ input\_str = ""\\
\small    \ \ \ \ while not input\_str:\\
\small    \ \ \ \ \ \ \ \ input\_str = input().strip()\\
\small    \ \ \ \ return input\_str\\
\small    \\
\small    line = read\_line()\\
\small    n, m = map(int, line.split())\\
\small    if not (0 <= n < 2000 and 0 <= m < 2000):\\
\small    \ \ \ \ sys.stderr.write("error at the boundary of n or m")\\
\small    \ \ \ \ exit()\\
\small    \\
\small    grid = [[0] * (m + 1) for \_ in range(n + 1)]\\
\small    for i in range(n + 1):\\
\small    \ \ \ \ line = read\_line()\\
\small    \ \ \ \ if len(line) != m + 1:\\
\small    \ \ \ \ \ \ \ \ sys.stderr.write("error at the boundary of the grid")\\
\small    \ \ \ \ \ \ \ \ exit()\\
\small    \ \ \ \ for j in range(m + 1):\\
\small    \ \ \ \ \ \ \ \ if not (0 <= int(line[j]) <= 1):\\
\small    \ \ \ \ \ \ \ \ \ \ \ \ sys.stderr.write("error at the boundary of the grid")\\
\small    \ \ \ \ \ \ \ \ \ \ \ \ exit()\\
\small    \ \ \ \ \ \ \ \ grid[i][j] = int(line[j])\\
\small    \\
\small    result = 0\\
\small    for i in range(n + 1):\\
\small    \ \ \ \ for j in range(m + 1):\\
\small    \ \ \ \ \ \ \ \ if grid[i][j] == 1:\\
\small    \ \ \ \ \ \ \ \ \ \ \ \ continue\\
\small    \ \ \ \ \ \ \ \ minDistance = math.inf\\
\small    \ \ \ \ \ \ \ \ for x in range(n + 1):\\
\small    \ \ \ \ \ \ \ \ \ \ \ \ for y in range(m + 1):\\
\small    \ \ \ \ \ \ \ \ \ \ \ \ \ \ \ \ if grid[x][y] == 1:\\
\small    \ \ \ \ \ \ \ \ \ \ \ \ \ \ \ \ \ \ \ \ \ distance = (x - i) ** 2 + (y - j) ** 2\\
\small    \ \ \ \ \ \ \ \ \ \ \ \ \ \ \ \ \ \ \ \ \ minDistance = min(minDistance, distance)\\
\small    \ \ \ \ \ \ \ \ result += minDistance\\
\small    \\
\small    print(result)\\
\thickhline
\end{tabularx}
\caption{
    {A real-life example of generated code for \texttt{CodeContests/2 (1575\_M. Managing Telephone Poles)}
    by \ours.}
}   
\end{table*}

\begin{table*}[t]

\begin{tabularx}{\textwidth}{X}
\thickhline
\textbf{Generated Test Cases}\\
\hdashline
\end{tabularx}
\renewcommand{\arraystretch}{0.7}
\begin{tabularx}{\textwidth}{X}
\footnotesize  \# Test Cases Regarding Functional Requirements\\
\footnotesize  \#\# General Cases\\
\footnotesize  \texttt{ERROR\_MESSAGE = "general case error"}\\
\footnotesize  \texttt{INPUT = "2 2\textbackslash n101\textbackslash n000\textbackslash n000"}\\
\footnotesize  \texttt{OUTPUT = "18"}\\
\footnotesize  \\
\footnotesize  \texttt{ERROR\_MESSAGE = "general case error"}\\
\footnotesize  \texttt{INPUT = "5 4\textbackslash n10010\textbackslash n00000\textbackslash n01000\textbackslash n00001\textbackslash n00100\textbackslash n00010"}\\
\footnotesize  \texttt{OUTPUT = "36"}\\
\footnotesize  \#\# Edge Cases\\
\footnotesize  \texttt{ERROR\_MESSAGE = "error at the lower boundary of n"}\\
\footnotesize  \texttt{INPUT = "0 2\textbackslash n00\textbackslash n00\textbackslash n00"}\\
\footnotesize  \texttt{OUTPUT = "0"}\\
\footnotesize  \\
\footnotesize  \texttt{ERROR\_MESSAGE = "error at the upper boundary of n"}\\
\footnotesize  \texttt{INPUT = "1999 2\textbackslash n" + "0" * 2000 + "\textbackslash n" + "0" * 2000}\\
\footnotesize  \texttt{OUTPUT = "0"}\\
\footnotesize  \\
\footnotesize  \texttt{ERROR\_MESSAGE = "error at the lower boundary of m"}\\
\footnotesize  \texttt{INPUT = "2 0\textbackslash n0\textbackslash n0\textbackslash n0"}\\
\footnotesize  \texttt{OUTPUT = "0"}\\
\footnotesize  \\
\footnotesize  \texttt{ERROR\_MESSAGE = "error at the upper boundary of m"}\\
\footnotesize  \texttt{INPUT = "2 1999\textbackslash n" + "0\textbackslash n" * 2000}\\
\footnotesize  \texttt{OUTPUT = "0"}\\
\footnotesize  \\
\footnotesize  \texttt{ERROR\_MESSAGE = "error when a\_\{i, j\} is not in the correct format"}\\
\footnotesize  \texttt{INPUT = "2 2\textbackslash n101\textbackslash n000\textbackslash n00"}\\
\footnotesize  \texttt{STD\_ERROR = "a\_\{i, j\} is not in the correct format"}\\
\footnotesize  \\
\footnotesize  \texttt{ERROR\_MESSAGE = "error when a\_\{i, j\} is not in the correct range"}\\
\footnotesize  \texttt{INPUT = "2 2\textbackslash n101\textbackslash n000\textbackslash n002"}\\
\footnotesize  \texttt{STD\_ERROR = "a\_\{i, j\} is not in the correct range"}\\
\footnotesize  \\
\footnotesize  \texttt{ERROR\_MESSAGE = "error when there is no telephone pole in the given grid"}\\
\footnotesize  \texttt{INPUT = "2 2\textbackslash n000\textbackslash n000\textbackslash n000"}\\
\footnotesize  \texttt{STD\_ERROR = "there is no telephone pole in the given grid"}\\
\footnotesize  \\
\footnotesize  \# Test Cases Regarding Non-functional Requirements\\
\footnotesize  \#\# Performance Requirements\\
\footnotesize  \texttt{ERROR\_MESSAGE = "execution failed while running exceptionally large input"}\\
\footnotesize  \texttt{INPUT = "1999 1999\textbackslash n" + "1" * 2000 + "\textbackslash n" * 1999 + "1" * 2000}\\
\footnotesize  \texttt{OUTPUT = "0"}\\
\footnotesize  \#\# Specific Quality Requirements\\
\footnotesize  \#\#\# Robustness\\
\footnotesize  \texttt{ERROR\_MESSAGE = "error when redundant spaces or newlines are added to the input"}\\
\footnotesize  \texttt{INPUT = "2 2\textbackslash n101\textbackslash n000\textbackslash n\textbackslash n\textbackslash n"}\\
\footnotesize  \texttt{OUTPUT = "18"}\\
\footnotesize  \\
\footnotesize  \texttt{ERROR\_MESSAGE = "error when the input is not in the correct format"}\\
\footnotesize  \texttt{INPUT = "2 2\textbackslash n101\textbackslash n000\textbackslash n00"}\\
\footnotesize  \texttt{STD\_ERROR = "the input is not in the correct format"}\\
\footnotesize  \\
\footnotesize  \texttt{ERROR\_MESSAGE = "error when the data type of the input is not correct"}\\
\footnotesize  \texttt{INPUT = "2 2\textbackslash n101\textbackslash n000\textbackslash n00a"}\\
\footnotesize  \texttt{STD\_ERROR = "the data type of the input is not correct"}\\
\footnotesize  \\
\footnotesize  \texttt{ERROR\_MESSAGE = "error when the input is not in the correct range"}\\
\footnotesize  \texttt{INPUT = "2 2\textbackslash n101\textbackslash n000\textbackslash n003"}\\
\footnotesize  \texttt{STD\_ERROR = "the input is not in the correct range"}\\
\footnotesize  \\
\footnotesize  \texttt{ERROR\_MESSAGE = "error when there is no telephone pole in the given grid"}\\
\footnotesize  \texttt{INPUT = "2 2\textbackslash n000\textbackslash n000\textbackslash n000"}\\
\footnotesize  \texttt{STD\_ERROR = "there is no telephone pole in the given grid"}\\
\footnotesize  \#\# Reliability\\
\footnotesize  Satisfied if no errors occur across all test cases\\
\footnotesize  \#\# Maintainability\\
\footnotesize  \texttt{ERROR\_MESSAGE = "error when cyclomatic complexity is more than the limit"}\\
\footnotesize  \texttt{COMPLEXITY\_LIMIT = 10}\\
\thickhline
\end{tabularx}
\caption{
    {A real-life example of generated test cases for \texttt{CodeContests/2/1575\_M. Managing Telephone Poles} by \ours. \textbf{\$\{Generated Code\}} denotes the string text of the code that is to be checked.}
}   
\end{table*}

\end{document}